\documentclass{aa}  

\usepackage{graphics,graphicx}
\usepackage{xcolor}
\usepackage{enumitem}

\usepackage{txfonts}
\usepackage{longtable}
\usepackage{lipsum}
\usepackage{hyperref}
\hypersetup{
  colorlinks=true,
  allcolors=[rgb]{0,0,0.8},
  pdftitle={The SRG/eROSITA all-sky survey: Hard X-ray selected Active Galactic Nuclei},
  pdfauthor={Waddell et al.},
  }

\newcommand{\ero}{eROSITA}

\newcommand{\fluxcgs}{~ergs~s$^{-1}$~cm$^{-2}$}
\newcommand{\lumcgs}{~ergs~s$^{-1}$}

\newcommand{\kev}{{\rm\thinspace keV}}
\newcommand{\ev}{{\rm\thinspace eV}}

\newcommand{\et}{{et al.\ }}

\def\cm{{\rm\thinspace cm}}
\def\pscm{\hbox{$\cm^{-2}\,$}}

\begin{document} 

   \title{The SRG/eROSITA all-sky survey: Hard X-ray selected Active Galactic Nuclei}

   \author{S. G. H. Waddell\inst{1}, 
          J. Buchner\inst{1},
          K. Nandra\inst{1},
          M. Salvato\inst{1},
          A. Merloni\inst{1},
          I. Gauger\inst{1},
          Th. Boller\inst{1},
          R. Seppi\inst{1, 2},
          J. Wolf\inst{1, 3, 4},
          T. Liu\inst{1},
          M. Brusa\inst{5, 6},
          J. Comparat\inst{1},
          T. Dwelly\inst{1},
          Z. Igo\inst{1,4},
          B. Musiimenta\inst{5, 6}
          }

   \institute{Max-Planck-Institut f\"{u}r extraterrestrische Physik, Giessenbachstrasse 1, 85748 Garching, Germany\\
              \email{swaddell@mpe.mpg.de}
        \and
            Department of Astronomy, University of Geneva, Ch. d’Ecogia 16, CH-1290 Versoix, Switzerland
        \and
             Exzellenzcluster ORIGINS, Boltzmannstr. 2, 85748, Garching, Germany
        \and
            Max-Planck Institut für Astronomie, Königstuhl 17, 69177 Heidelberg, Germany
        \and
            Dipartimento di Fisica e Astronomia "Augusto Righi", Alma Mater Studiorum Università di Bologna, via Gobetti 93/2, 40129 Bologna, Italy
        \and
            INAF-Osservatorio di Astrofisica e Scienza dello Spazio di Bologna, via Gobetti 93/3, 40129 Bologna, Italy           
        }

   \date{Received XXX; accepted YYY}

 
  \abstract
   {The eROSITA instrument aboard the Spectrum Roentgen Gamma (SRG) satellite has performed its first all-sky survey between December 2019 and June 2020. This paper presents the resulting hard X-ray ($2.3-5\kev$) sample, the first created from an all-sky imaging survey in the $2-8\kev$ band, for sources within western galactic sky (eROSITA\_DE). }
   {We produce a large uniform sample of hard-X-ray selected AGN, and to characterise them with supporting multi-wavelength astrometry, photometry and spectroscopy. For the 2863 sources within the sky coverage of the Legacy Survey DR10, counterparts are identified and classified. We also perform comparisons with the Swift BAT sample and HEAO-1 AGN sample to attempt to better understand the effectiveness and sensitivity of eROSITA in the hard band. }
   {The 5466 hard X-ray selected sources detected with eROSITA are presented and discussed. The Bayesian statistics-based code \texttt{NWAY} is used to identify the counterparts for the X-ray sources. These sources are classified based on their multiwavelength properties, and the literature is searched to identify spectroscopic redshifts, which further inform the source classification. A total of 2547 sources are found to have good-quality counterparts, and 111 of these are detected only in the hard band. }
   {Comparing with other hard X-ray selected surveys, the eROSITA hard sample covers a larger redshift range and probes dimmer sources, providing a complementary and expanded sample as compared to Swift-BAT. Examining the column density distribution of missed and detected eROSITA sources present in the follow-up catalog of Swift BAT 70 month sources, it is demonstrated that eROSITA can detect obscured sources with column densities $>10^{24}$\pscm, but that the completeness drops rapidly after $10^{23}$\pscm. A sample of hard-only sources, many of which are likely to be heavily obscured AGN, is also presented and discussed. X-ray spectral fitting reveals that these sources have extremely faint soft X-ray emission and their optical images suggest that they are found in more edge-on galaxies with lower b/a.  }
   {The first eROSITA all-sky survey has provided the first imaging survey above $2\kev$, and the resulting X-ray catalog is demonstrated to be a powerful tool for understanding AGN, in particular heavily obscured AGN found in the hard-only sample. }

    \titlerunning{Hard X-ray selected AGN in eRASS1}
    \authorrunning{Waddell et al.}
   \maketitle
%

\section{Introduction}

The primary science driver of eROSITA \citep[extended Roentgen Survey with an Imaging Telescope Array;][]{Merloni2024} onboard the Spectrum Roentgen Gamma \citep[SRG;][]{Sunyaevero} mission is to detect X-ray emission from $\sim10^5$ clusters of galaxies, enabling X-ray mapping of the large scale structure of the Universe \citep{Merloniero, Predehlero}. However, by far the most prevalent sources of X-ray emission detected by eROSITA will originate in Active Galactic Nuclei (AGN), driven by powerful supermassive black holes which are actively accreting material \citep[see for example,][Nandra \et submitted]{2022Salvato}. AGN emit across all wavelengths of electromagnetic radiation, from radio to X-ray and $\gamma$-rays, and may even outshine their host galaxies \citep[see e.g.][for a review]{2017Padovani}. By studying these sources in the X-rays, these effects may be investigated, as the X-ray emission originates closest to the supermassive black hole \citep[e.g. see][]{1978Elvis}. The primary source of these X-rays is thought to be the corona \citep[e.g.][]{1991Haardt, 1993Haardt, 2000Merloni, 2004Fabian}, a population of relativistic electrons which Compton up-scatters UV photons from the accretion disc. The result is an X-ray spectrum which takes the approximate form of a power law, with a low-energy turn around of $\sim1\ev$ \citep{1973discmod} corresponding to the temperature of the UV photons in the inner accretion disc, and a high-energy turn-over at $100-300\kev$ \citep[e.g.][]{2001Petrucci,2015Fabian,2018Tortosa,2019Middei} corresponding to the electron temperature in the corona. 

The AGN unification model \citep{1993Antonucci, 1995Urry} has as an interpretation for two observationally observed two subgroups \citep{1943Seyfert,1974seyferts,1985Osterbrock}; Seyfert 1 (type-1) AGN are unobscured, revealing broad optical spectral lines and typically showing strong soft X-ray emission, while Seyfert 2 (type-2) AGN are obscured by a cold, dusty distribution of clouds referred to as the torus. The optical spectra therefore do not show evidence for broad lines, and often, the X-ray spectra reveal weakened soft X-ray emission due to obscuration with column densities of the order of N$_{\rm H}$ $>10^{22}$\pscm \citep{2014Merloni,2016Burtscher}. When the absorbing column density is sufficiently high ($\sim10^{22.5-23}$\pscm), the source will lack observable soft X-rays, only revealed at energies above $\sim2\kev$. When the column density exceeds the inverse of the Thomson cross-section ($\sigma_T^{-1} \simeq 1.5\times10^{24}$\pscm), then the source is defined as Compton thick and the coronal emission is largely obscured below $10\kev$ \citep[e.g.][]{2004Comastri}. Given that a large fraction or even majority of AGN in the local Universe are believed to be obscured \citep[e.g.][]{1995Maiolino,1998Maiolino,2000Matt,2009Fiore}, and that a correspondingly large portion of the accretion energy density of the Universe is believed to be contained in obscured AGN \citep[e.g.][]{1999Fabian,1999FI,2015Aird,2015Buchner}, detecting these sources is of particular importance. 

Numerous surveys making use of X-ray telescopes with sensitivity above $2\kev$ have been conducted to detect obscured AGN. First, the non-imaging telescope HEAO-1 \citep{1979Rothschild} surveyed the entire sky in the $0.2-25\kev$ band, creating a very high-flux sample of hard X-ray AGN presented in \citet{1982Piccinotti}, which have been re-examined in many later works. Later works have used imaging telescopes covering smaller and/or non-contiguous areas of the sky; for example, using ASCA \citep{1999Ueda}, BeppoSAX \citep{2008Della}, Chandra \citep[e.g.][]{2015Nandra} and XMM-Newton \citep[e.g.][]{2003Fiore,2007Brusa,2007Cocchia,2007Hasinger}. These missions are mostly sensitive below $10\kev$; complementary, higher energy missions, including NuSTAR \citep[$3-79\kev$;][]{2013nustar} and Swift BAT \citep[$14-195\kev$;][]{2004Gehrels}, have been employed to detected heavily obscured, Compton-thick AGN. Swift BAT has performed an all-sky survey in the ultra-hard X-ray band, with the most recent 105 month release \citep{2018Oh} detecting 1632 X-ray sources, including at least 1105 AGN. Many of these sources have been followed up with NuSTAR, whose improved energy sensitivity and softer X-ray coverage has enabled spectral modelling to compare torus models \citep{2009Ikeda,2009Murphy,2011Brightman,2018borus,2019borus,2019Buchner}. 

As a precursor to the all-sky surveys, the eROSITA Final Equatorial Depth Survey \citep[eFEDS;][]{Brunnerefeds} provided key insights as to the science capabilites of eROSITA surveys. Before the eROSITA all-sky survey, eFEDS was at the time the largest contiguous dedicated X-ray imaging survey. eFEDS is designed to mimic the depth of the full planned, 8-pass all-sky survey of eROSITA,  From the 144 square-degree survey, a small hard X-ray selected sample of 246 sources was identified. These sources are classified and analysed in Nandra \et (submitted) and \citet{2023Waddell}, as well as by \citet{2022Brusa}. From this analysis, a sample of 200 AGN were identified. It was found that the AGN covered a wide luminosity range, and had spectroscopic redshifts up to $z\sim3$, with a peak in the distribution at $z\sim0.3$, suggesting that we primarily selected a local sample of bright AGN. By selecting sources which did not appear to be present in the soft band ($0.2-2.3\kev$), a small sample of highly obscured AGN were recovered, with column densities up to $\log N_{\rm H}>23$. Other AGN, particularly some of the brighter sources at higher redshifts, were found to be blazars, which should be treated with care. Through spectral analysis, \citet{2023Waddell} also showed that many sources require additional spectral components beyond a simple obscured power law, including a warm absorber \citep[e.g.][]{1998George,2000Kaspi,2004Blustin,2005Blustin,2004Gierlinski,2019Mizumoto}, a neutral partial covering absorber \citep[e.g.][]{2004Tanaka,2021Parker}, or a soft excesses \citep[e.g.][]{1981Pravdo,1985Arnaud,1985Singh,1999Ross,2005Ross,2012Done,2018Petrucci,2020Petrucci,2023Waddell} to fully model the spectral complexity. The eFEDS hard sample thus illustrated potential for detailed AGN analysis with the full all-sky survey.

The eRASS1 hard X-ray selected sample is highly complementary to these works. Covering half the sky, the vast size of the survey has yielded a large sample of hard X-ray selected sources \citep{Merloni2024}. Although eROSITA lacks sensitivity above $8\kev$, it is highly sensitive to soft X-ray emission, enabling accurate modelling of the obscuring column densities. Furthermore, with excellent soft X-ray spectral resolution, eROSITA spectra can be fit with physically motivated models to better understand the nature of X-ray emission and absorption components \citep{2023Waddell}. Selecting hard-X-ray detected sources which lack soft X-ray emission can also allow for the identification of obscured AGN.

In this paper, we present the counterparts, classifications and an analysis of the X-ray and optical properties of the 5466 hard X-ray selected sources in the first all-sky survey. We focus on X-ray point sources which have not been flagged as potentially spurious or problematic, with a particular focus on the identification of AGN. In Section 2, we provide an expansion of the X-ray properties of the catalog described in \citet{Merloni2024}, and define a hard-only sample with no significant detection of emission below $2.3\kev$. In Section 3, information on the counterparts within the DESI imaging legacy surveys data release 10 \citep{2019Dey} is described, spectroscopic redshifts are compiled, where available, sources are classified, and beamed AGN are flagged. In Section 4, we briefly describe AGN matched from outside the Legacy survey. Section 5 presents a comparison to the Swift BAT survey and section 6 compares to the HEAO-1 AGN sample. Section 7 briefly discusses variable sources, and in particular, variable AGN. The spectroscopic AGN sample is presented in Section 8, with a particular focus on hard-only AGN. Further discussion is given in Section 9, and conclusions are given in Section 10.

Throughout this paper, we adopt a $\Lambda$CDM cosmology with $\Omega_{\Lambda} =0.7$, $\Omega_m =0.3$, and $H_0 = 70$ km s$^{-1}$ Mpc$^{-1}$. 

\section{Properties of the eRASS1 hard sample}
\subsection{Population information}

\begin{figure*}
    \centering
    \includegraphics[width=0.85\textwidth]{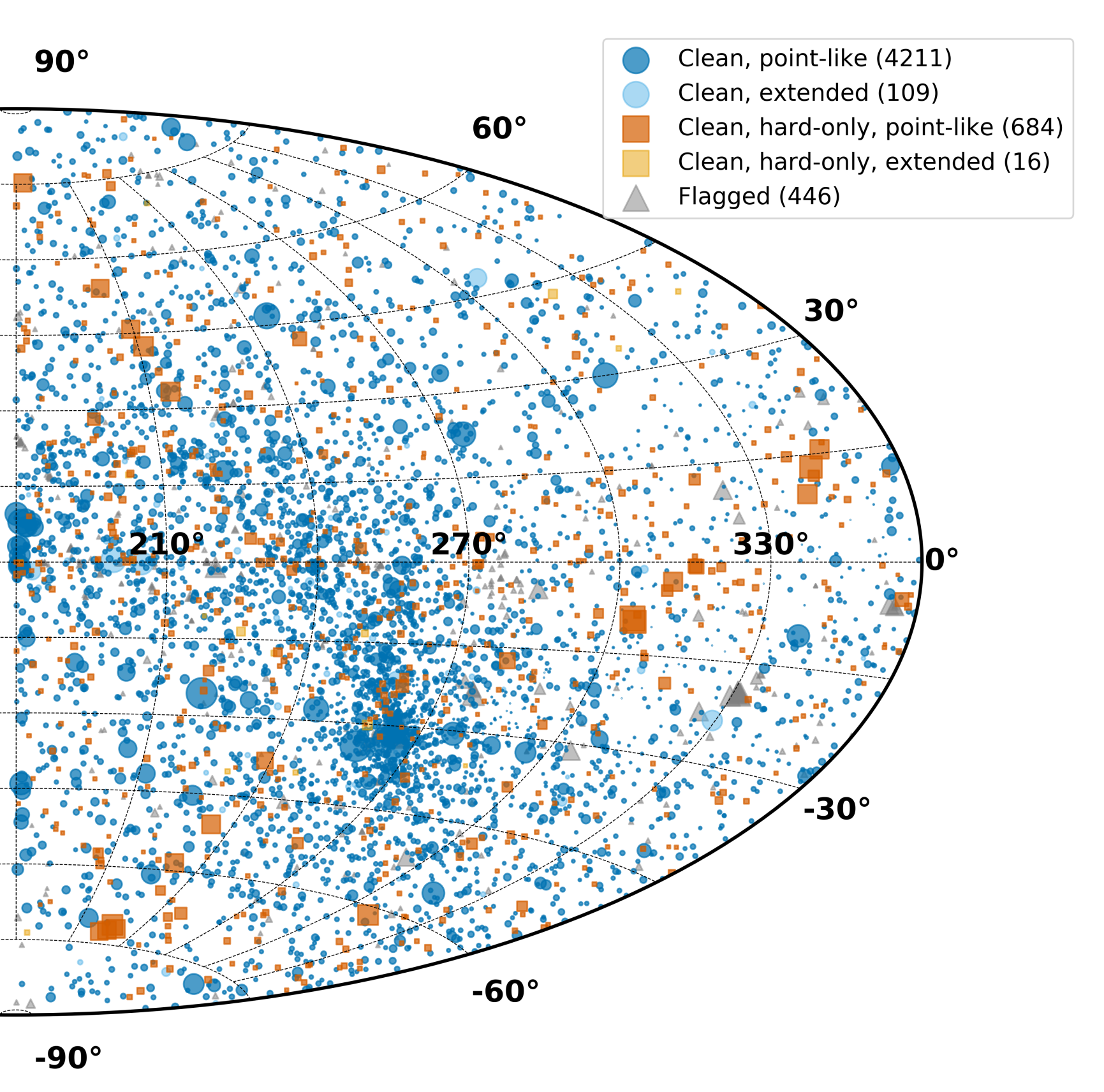}
    \caption{Hard X-ray detected sources in the first all-sky survey. The non-flagged, point-like sample is shown as blue circles, and non-flagged extended sources are shown as pale blue circles. The non-flagged, point-like, hard-only sample is shown with orange squares, and the hard-only extended sources are shown with light orange squares. Sources flagged as problematic are shown as grey triangles. The size of the shapes corresponds to the hard band ($2.3-5\kev$) flux, with larger shapes corresponding to brighter X-ray fluxes. The values total 5466 sources. }
    \label{fig:radec}
\end{figure*}

The hard ($2.3-5\kev$) X-ray selected catalog from the eRASS1 all sky survey in the western galactic sky (eROSITA\_DE) is presented in \citet{Merloni2024}, and the X-ray positions of all sources are shown in Fig~\ref{fig:radec}. Relevant additional sub-samples are also shown in this plot, which will be described in the following sub-sections. The catalog contains a total of 5466 sources, with a hard-band detection likelihoods of \texttt{DET\_LIKE\_3}$>12$, where band 3 is $0.6-2.3\kev$ \citep{Merloni2024}. Based on simulations performed in \citet{2022Seppi}, this indicates a spurious fraction of 10\% based on all-sky estimates. Out of the 5466 sources, 5087 sources are considered to be point-like in the X-ray, having a EXT\_LIKE (extended likelihood) of 0. The other 379 sources are candidate extended sources in the X-ray. Most of these will be analysed in more detail in Bulbul \et (in press) and/or Bahar \et (submitted), as many are likely cluster of galaxies, while others are likely extended galactic sources including supernova remnants. These are discussed in more detail in Section 2.3. 

A number of flags indicating spurious or otherwise problematic sources are presented in the catalog, and are outlined in Table 5 of \citet{Merloni2024}. These include five flags for potentially spurious sources, identified due to their proximity to SNR, bright point sources, stellar clusters, large local galaxies, known galaxy clusters, or sources potentially contaminated with optical loading. A further three flags identify sources for which no error bars were provided on the coordinate errors, extent errors, or source count errors. While these sources are not neccessarily spurious, they are removed for analysis as these parameters are necessary for sample selection and counterpart identification, and cannot be trusted for flagged sources. When these flagged sources are removed, a total of 5020 sources remain, 4895 of which are point-like. 

\begin{figure}
    \centering
    \includegraphics[width=0.95\columnwidth]{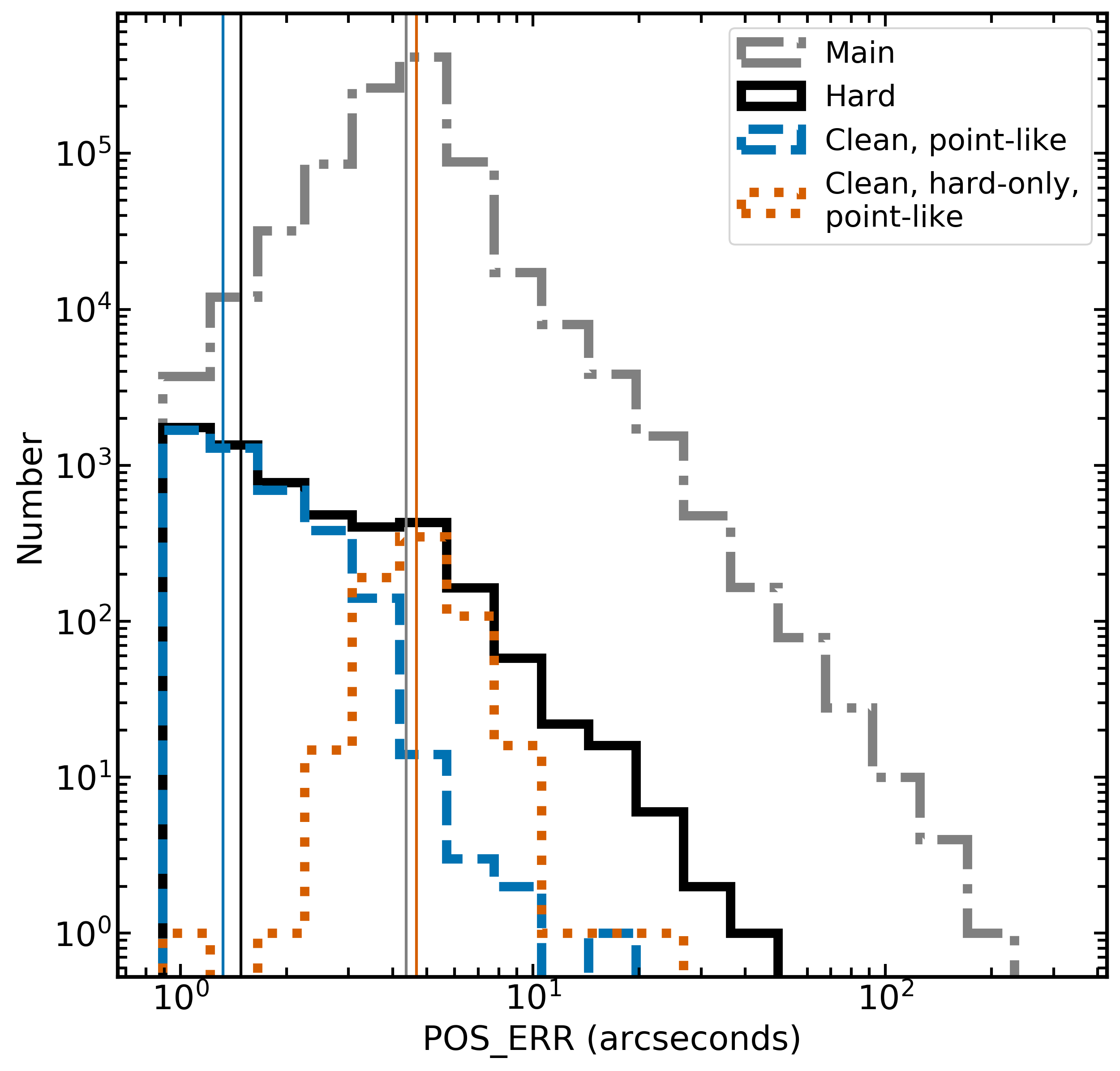}
    \caption{Distribution of errors on the coordinates (\texttt{POS\_ERR}) for individual sources. The eRASS1 main sample is shown with a grey, dash-dot line. The hard sample is shown as a black solid line, and is further sub-divided into non-flagged, point-like hard sources, and point-like hard-only sources, in a blue dashed line and an orange dotted line, respectively. Median values are indicated with vertical lines in corresponding colours. }
    \label{fig:radecerr}
\end{figure}

A key difference between the main and hard X-ray samples is the distribution of the positional uncertainty, contained in the column \texttt{POS\_ERR}. These can be seen in Fig.~\ref{fig:radecerr}, where sources in the main sample are shown in grey, sources in this hard sample are shown in black, hard-only sources are shown in orange, and sources in the hard sample with soft detections are shown in blue. The astrometric calibration \citep{Merloni2024} imposes a lower bound of 0.9" in the calculation of \texttt{POS\_ERR}. In the main sample, the peak of the distribution is found at $\sim4.5"$, whereas for the hard sample it is found around $1.5"$. This is because both the soft and hard X-ray positional information is used to localize the sources, reducing the positional errors. This also affects the counterpart identification procedure, since preference will be given to sources within the positional error. There is also a difference observed for the hard-only sources; since their positions are only determined by the hard X-rays, the positional errors are larger and the distribution more closely follows that of the main sample. Many sources flagged as spurious also have large or absent positional errors, accounting for the large number of sources in the black histogram above $1"$. 

The distribution of X-ray counts and a comparison to the main sample can be found in Figure 5 of \citet{Merloni2024}. The distribution peaks at only $\sim7-8$ counts in the hard band, but extends up to tens of thousands for the brightest sources. More detailed flux distributions will be presented in Section 3 of this work.
\subsection{Hard-only sources}
\begin{figure}
    \centering
    \includegraphics[width=0.95\columnwidth]{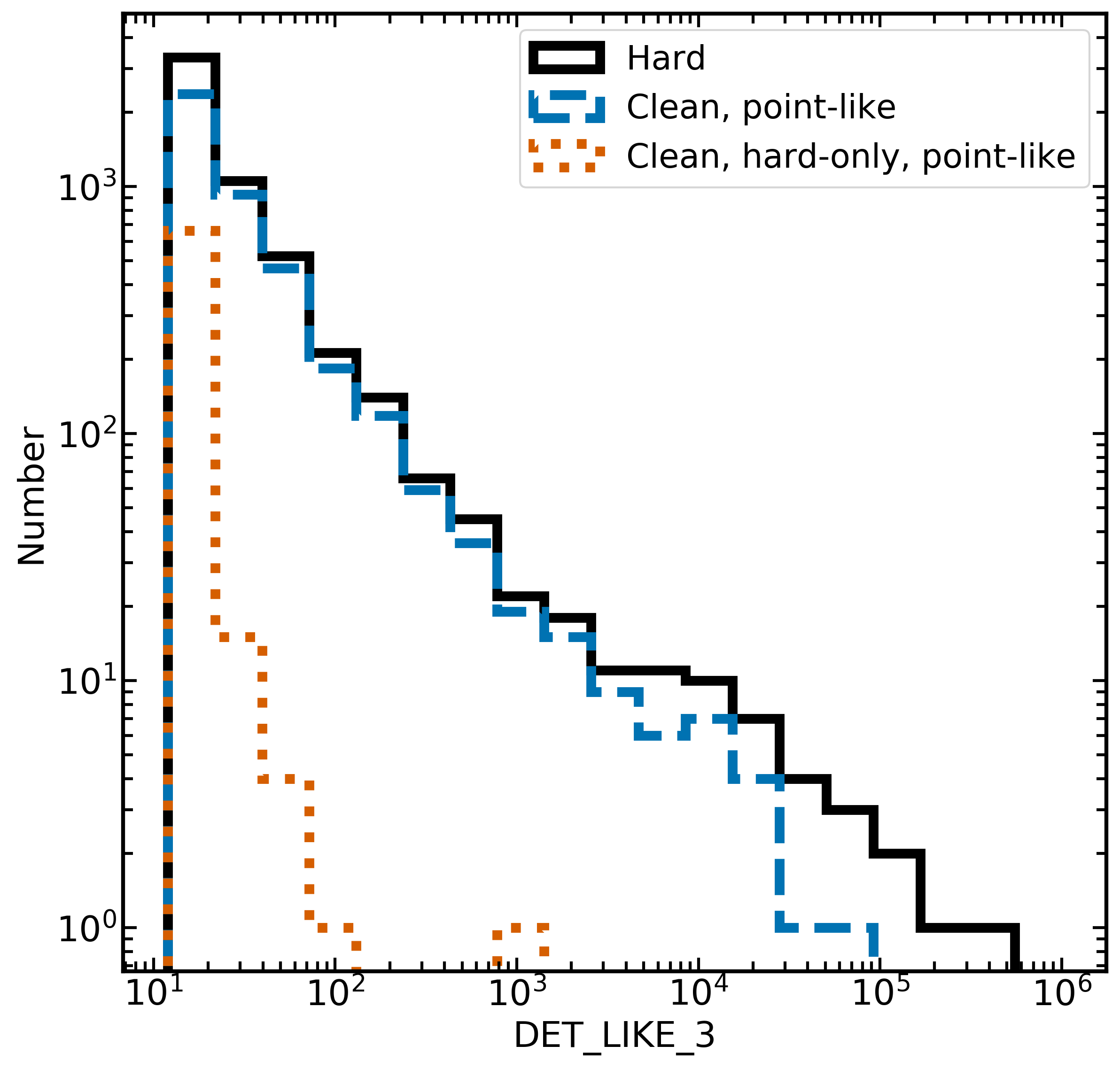}
    \caption{Distribution of \texttt{DET\_LIKE\_3} values. The hard sample is shown as a black solid line, and is further sub-divided into point-like sources, and point-like hard-only sources, in a blue dashed line and an orange dotted line, respectively.}
    \label{fig:d3}
\end{figure}

In this work, hard-only sources are defined as those having:

\begin{equation}
    \rm{DET\_LIKE\_1}<5 \hspace{1.5mm} \rm{\&} \hspace{1.5mm} \rm{DET\_LIKE\_2}<5
\end{equation}

\noindent where 1 indicates the $0.2-0.6\kev$ band and 2 indicates the $0.6-2.3\kev$ band. In this way the detection likelihood in neither band is equal to the supplementary sample cut-off, at \texttt{DET\_LIKE\_0}>5 (where 0 indicates the $0.2-2.3\kev$ band). In total, 738 sources meet this criteria. These sources are distributed across the sky, with many being in or near the galactic plane, as these X-ray sources would be obscured by the Milky Way disc. They have higher positional errors than other sources in the hard sample, complicating counterpart identification. Fig.~\ref{fig:d3} shows the distribution of \texttt{DET\_LIKE\_3}, the detection likelihood in the hard band. In general, the hard-only sources have lower \texttt{DET\_LIKE\_3} values; this is sensible, as eROSITA is extremely sensitive in the soft band and should easily be able to detect soft X-ray emission, if present; so any hard-only sources must be very dim and/or heavily obscured. There is also likely to be a very high spurious fraction among the hard-only sample; given that $\sim10\%$ of sources are expected to be spurious, $\sim550$ hard-only sources may be spurious, leaving only $\sim200$ real sources. Identifying these is of particular interest, as these are important objects for future study.

\begin{figure}
    \centering
    \includegraphics[width=0.95\columnwidth]{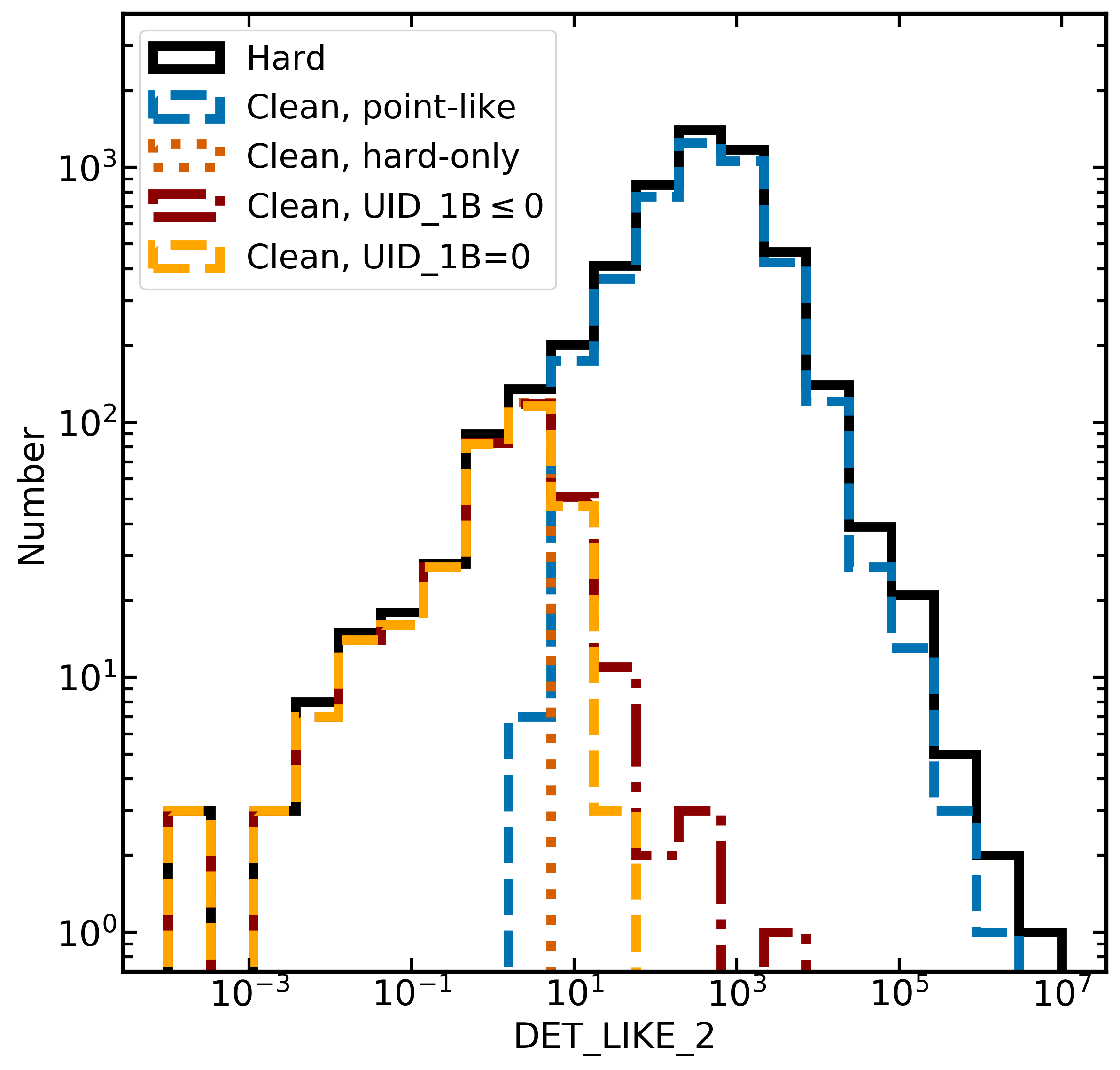}
    \caption{Distribution of \texttt{DET\_LIKE\_2} values. The hard sample is shown as a black solid line, and is further sub-divided into non-flagged, point-like hard sources, and point-like hard-only sources, in a blue dashed line and an orange dotted line, respectively. The hard-only samples as defined in \citet{Merloni2024} are also shown, with the higher fidelity, \texttt{UID\_1B}=0 sample shown as a pale orange dashed line, and the less pure, \texttt{UID\_1B}$\leq0$ sample shown as a dark red, dash-dot line. }
    \label{fig:d2}
\end{figure}

Examining only the point-like hard-only sources, there are 684 meeting the above definition (Eq. 1). Our definition differs from that which is presented in \citet{Merloni2024}, which relies on spatial coincidence of sources in the main and hard catalogs (weak match) and a ratio of their counts (strong match). Sources which are weak matches have a negative value in the \texttt{UID\_1B} column in the hard X-ray catalog, and those no matches have a 0 in the \texttt{UID\_1B} column. Selecting the point-like sources, 729 have no matches, and 752 have no or weak matches. However, examining the distribution of fluxes and detection likelihoods in band 2 ($0.6-2.3\kev$; see Fig.~\ref{fig:d2}), there are some sources with high detection likelihoods of tens to thousands. These sources are extremely unlikely to be true, hard-only sources. By using the modified definition of Eq. 1, all of these sources are removed. Out of the 684 hard-only sources from this work, 673 are also hard-only using the definition in \citet{Merloni2024}, demonstrating very high overlap between the two selections. 

Also of interest are the sources for which the softest band has a significant detection likelihood, but the median band does not; that is, having $\texttt{DET\_LIKE\_1}>5$ \hspace{1.5mm} \rm{\&} \hspace{1.5mm} $\texttt{DET\_LIKE\_2}<5$. There are five such sources, all with $\texttt{DET\_LIKE\_1}<9$. Four of these sources are in the LS10 area, and none have any flags in the X-ray. Such sources may be candidates for obscured AGN, with significant soft X-ray leakage due to Compton scattering or an ionised absorption medium.

\subsection{Extended sources}
There are a total of 379 sources in the hard extended sample, peaking at an extent likelihood of EXT\_LIKE $\sim140-150$. This is very different from the extent likelihood distribution of the main sample, which peaks at EXT\_LIKE $\sim3-4$. This is because the hard sample primarily selects the brightest sources in the eROSITA X-ray sky, facilitating the identification of the sources type. Of the 379 sources, many are flagged as lying near a known cluster or SNR, and are thus likely spurious. Removing these, a total of 125 sources remain. Of these, 16 have been identified as being hard-only sources which satisfy equation 1. The remaining sources are very likely found in the main catalog and will be described in Bulbul \et (in press) and Kluge \et (submitted). Of the 16 hard-only extended sources, one is positionally coincident with a source in the main catalog, and only three have sufficiently high extent likelihoods (EXT\_LIKE $>9$) to make it into the secure cluster catalog (Bulbul \et in press). Two sources are in the galactic plane, and the remaining source is positionally coincident with a known galaxy, LEDA 483209. For the remainder of this work, the focus will be restricted to the 5087 point-like X-ray sources.

\section{Counterparts in the legacy survey footprint}
\subsection{Counterpart identification}
To attempt to identify the counterparts of the eRASS1 Hard sample, the photometric and other data provided in the DESI imaging legacy surveys, specifically the Legacy Survey Data Release 10 \citep[LS10;][]{2019Dey} are used. In addition to g-, r-, i- and z- band photometry, LS10 also includes WISE fluxes, obtained from forced-flux photometry on the unWISE data at the positions of the Legacy Survey optical positions. LS10 is the first of the Legacy Survey data releases to include i-band photometry, and coverage is improved in the Southern Hemisphere, coincident with a large portion of the eROSITA-DE sky. In all, 15342 deg$^2$ of the sky are covered by at least one pass in all four photometry bands. This is defined using a multi-order coverage map (MOC), and is the sky area covered by all those 0.25x0.25deg legacy survey 'bricks' which have at least 1000 detections, and which have estimated 5 sigma point-source depths of at least g>23.5, r>23.3, i>22.8 and z>22.3, in AB magnitudes.Of the 4895 X-ray point-like sources without flags, 2863 are in the legacy survey area, defined as having coverage in all four photometric bands. Out of these, 363 are hard-only sources in the LS10 area. These numbers and other key values of interest are listed in Table~\ref{tab:nums}.

\renewcommand{\arraystretch}{1.2}
\begin{table}
\caption{Useful reference numbers for the different samples listed throughout this work.  }
\label{tab:nums}
\resizebox{0.47\textwidth}{!}{%
\centering
\begin{tabular}{ll}
\hline
Category  & No. of sources \\
\hline
Hard X-ray selected sample & 5466 \\
Point-like sources & 5087 \\
Clean, point-like sources (CPL) & 4895 \\
Hard-only CPL sources & 684 \\
CPL sources in the LS10 area & 2863 \\
Hard-only CPL in LS10 & 363 \\
CPL soft sources outside the LS10 area & 1710 \\
Hard-only outside LS10 & 321 \\
CPL in LS10 with good p\_any & 2547 \\
Hard-only good p\_any & 111 \\
Good p\_any with spec-z>0.002 & 1243\\
Hard-only good p\_any with spec-z>0.002 & 23 \\
Stars & 325 \\
Beamed AGN & 319 \\
BAT matches & 487 \\
CPL BAT matches & 456 \\
Other AGN outside LS10 & 692 \\
\hline
Final spec-z sample (Section 8.1) & 1328 \\
Final spec-z hard-only & 29 \\
\hline
\hline
\end{tabular}
}
\end{table}

Counterpart identification is done using \texttt{NWAY} \citep{2018Salvato,2022Salvato}, a Bayesian framework which has been specifically designed to identify counterparts for astronomical sources. This method uses positional information, positional errors, and the local X-ray source density as well as the multiwavelength properties to assess the likelihood of an optical source being an X-ray emitter as well as being the correct counterpart. The training is performed on a 4XMM sample cut at a $0.5-2\kev$ flux of $2\times10^{-15}$\fluxcgs, which corresponds to the minimum flux of the hard sample. More details of the training are described in \citet[][Salvato \et (in prep.)]{2022Salvato}, in which the same method and training priors are applied to the main eRASS1 catalog. Some caveats to this will be discussed in Section 9.1. 

Starting from the full hard X-ray catalog, the extended sources are removed. \texttt{NWAY} is then run on the sample to find the X-ray counterparts. LS10 sources within a 60 arcsecond radius of the X-ray source are considered, in order to account for the few X-ray sources with very large positional error (maximum $\sim50$ arcseconds), some of which are hard-only sources which are of particular interest in this work. Out of the 2863 point-like, non-problematic X-ray sources in the LS10 area, 2850 sources have a primary counterpart (match\_flag=1).

\subsection{Sources without a counterpart}
Since 2850 out of 2863 sources have at least one counterpart, this leaves a mere 13 sources without a LS10 counterpart. These sources are listed in Table~\ref{tab:noctp}. To attempt to identify these sources, we examine the sky in the legacy survey viewer tool \footnote{\url{https://www.legacysurvey.org/viewer}}, and query their positions in SIMBAD \citep{2000simbad} using a two arc minute match radius. This extremely large search radius allows us to check if there are any known sources in the general area of the X-ray detections, but this is not used to assign counterparts. Notes on the results of these searches can be found in the far right column of Table~\ref{tab:noctp}. 

In total, three sources are clearly associated with bright AGN which may have issues with their photometry in LS10, one source is coincident with a bright star, one source is coincident with a cluster of galaxies, one source appears in a gap in LS10 which is not reflected in the MOC, and one source has no previously named nearby sources. A further six sources are all found in the South Ecliptic Pole (SEP) area \citep{Merloni2024}. There are issues with the shallow depth of the LS10 photometry as well as the eROSITA fluxes in this region due to the repeated exposures, so it is reasonable that some sources are missing counterparts there. The \texttt{DET\_LIKE\_3} values are high for most sources, so it is likely that most are real X-ray sources. The three sources which have spectroscopic redshifts are manually added to the spec-z catalog in Section 8.1.

\renewcommand{\arraystretch}{1.3}
\begin{table*}
\caption{Non-flagged, point sources without counterparts in LS10. The eROSITA name is given in column (1), and the X-ray RA and Dec are given in columns (2) and (3). Column (4) gives the hard band detection likelihood (\texttt{DET\_LIKE\_3}), and column (5) lists notes for each source. Coincident indicates $<5$ arcsecond match radius, while nearby is defined as $<2$ arcminutes. }
\label{tab:noctp}
\resizebox{\textwidth}{!}{%
\begin{tabular}{lllll}
\hline
(1) & (2) & (3) & (4) & (5) \\
eROSITA Name & RA & Dec & \texttt{DET\_LIKE\_3} & Notes \\
\hline
1eRASS J015513.1-450612 & 28.8050 &  -45.1035 & 17.56 & Coincident with bright quasar QSO J0155-4506 (z=0.451)   \\
1eRASS J033336.3-360825 & 53.4013 &  -36.1403 & 969.98 & Coincident with Sy1 NGC 1365 (z=0.00554) \\  
1eRASS J041524.0-232234 & 63.8501 &  -23.3763 & 15.20 & Coincident with QSO [VV2006] J041524.0-232234 (z=0.6163)  \\
1eRASS J041619.1-480412 & 64.0796 &  -48.0700 & 18.69 & No known nearby objects in SIMBAD \\
1eRASS J041722.9-474838 & 64.3452 &  -47.8107 &  19.87 & Coincident with galaxy cluster SPT-CL J0417-4748  \\
1eRASS J053232,8-655141 & 83.1373 &  -65.8614 & 75.67 & Hole in Legacy Survey coverage \\
1eRASS J055909.1-663206 & 89.7878 &  -66.5351 &  20.27 & In SEP   \\
1eRASS J055927.9-662925 & 89.8662 &  -66.4904 &  45.51 & In SEP   \\
1eRASS J055934.2-653831 & 89.8927 &  -65.6421 & 17.59 & In SEP \\
1eRASS J060008.9-662505 & 90.0371 &  -66.4182 &  40.40 & In SEP  \\
1eRASS J060012.4-662731 & 90.0517 &  -66.4587 &  19.42 & In SEP  \\
1eRASS J060015.5-663231 & 90.0645 &  -66.5421 &  119.46 & In SEP  \\
1eRASS J185305.9-501051 & 283.2746 &  -50.1810 &  20.52 & Coincident with star V* PZ Tel   \\
\hline
\hline
\end{tabular}
}
\end{table*}

\subsection{Calibration of \texttt{p\_any} thresholds}
In order to assess the quality of the counterpart match, the \texttt{p\_any} values produced by \texttt{NWAY} are used. For each entry in the X-ray catalogue, the \texttt{p\_any} is defined as the probability that there is a counterpart. To assign thresholds to these values, a new X-ray catalog is produced, where the RA and Dec of the X-ray sources have each been shifted by two arcminutes. Sources at these new shifted positions which are coincident with an original X-ray source are then removed. In this way, it is possible to evaluate the incidence of chance alignment in the \texttt{NWAY} cross-matching procedure by comparing the p\_any distributions of real and spurious matches.

\begin{figure}
    \centering
    \includegraphics[width=0.95\columnwidth]{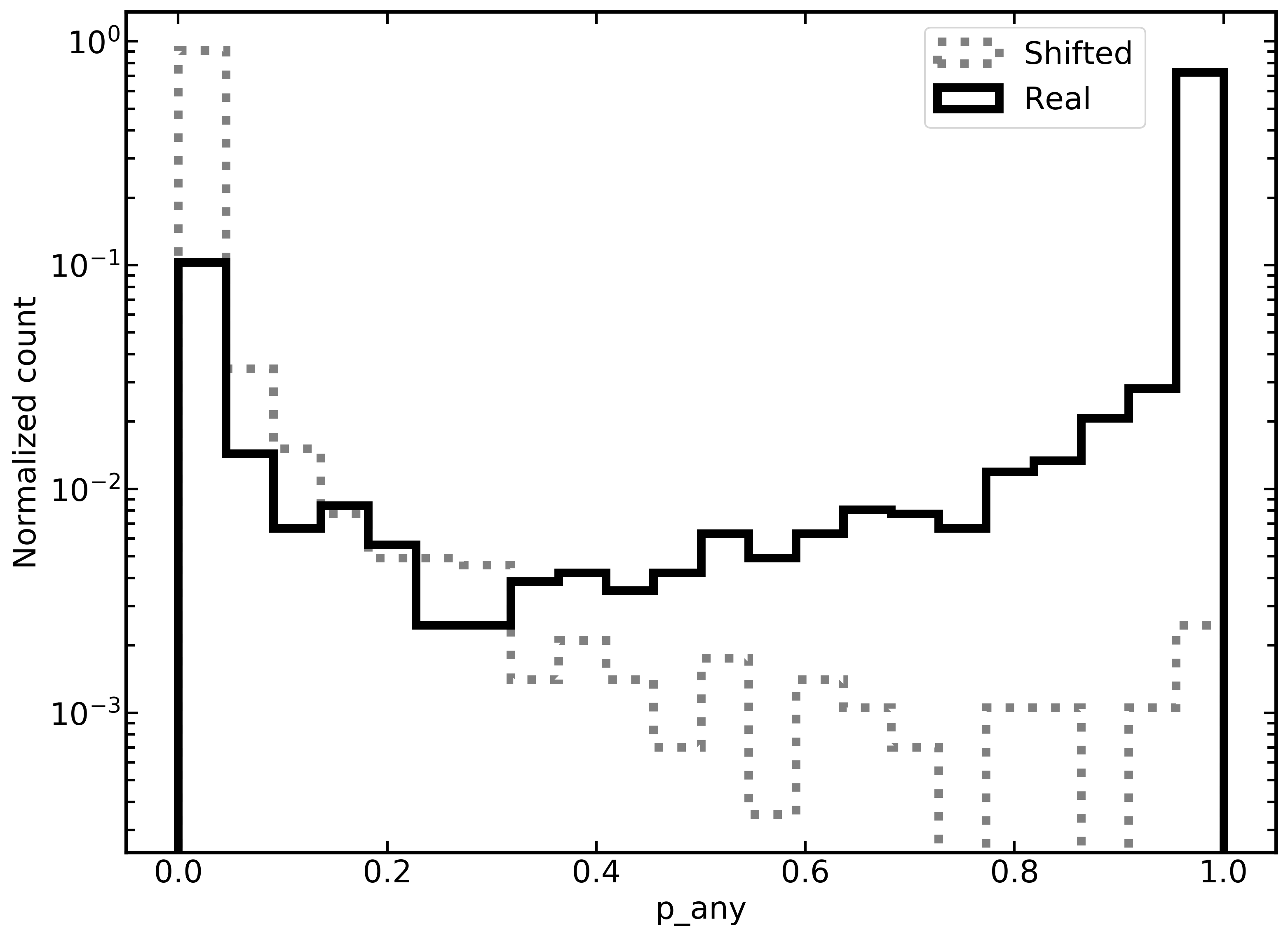}
    \caption{Distribution of \texttt{p\_any} thresholds for the real (black, solid line) and shifted (gray, dotted line) samples. }
    \label{fig:pany}
\end{figure}

The (normalized) distributions of the \texttt{p\_any} values obtained from the real and shifted samples are shown in Fig.~\ref{fig:pany}. Only point sources with no flags in the X-ray are considered. For the shifted sources, only a very small minority of sources have large \texttt{p\_any}, showing that \texttt{NWAY} rarely finds a secure counterpart where none should be found. There is a very high peak at $\sim0$, with most sources having very low \texttt{p\_any} values. In contrast, the \texttt{p\_any} values of the real sources are more broadly distributed, with the largest peak at a \texttt{p\_any} of 1. 

\begin{figure}
    \centering
    \includegraphics[width=0.95\columnwidth]{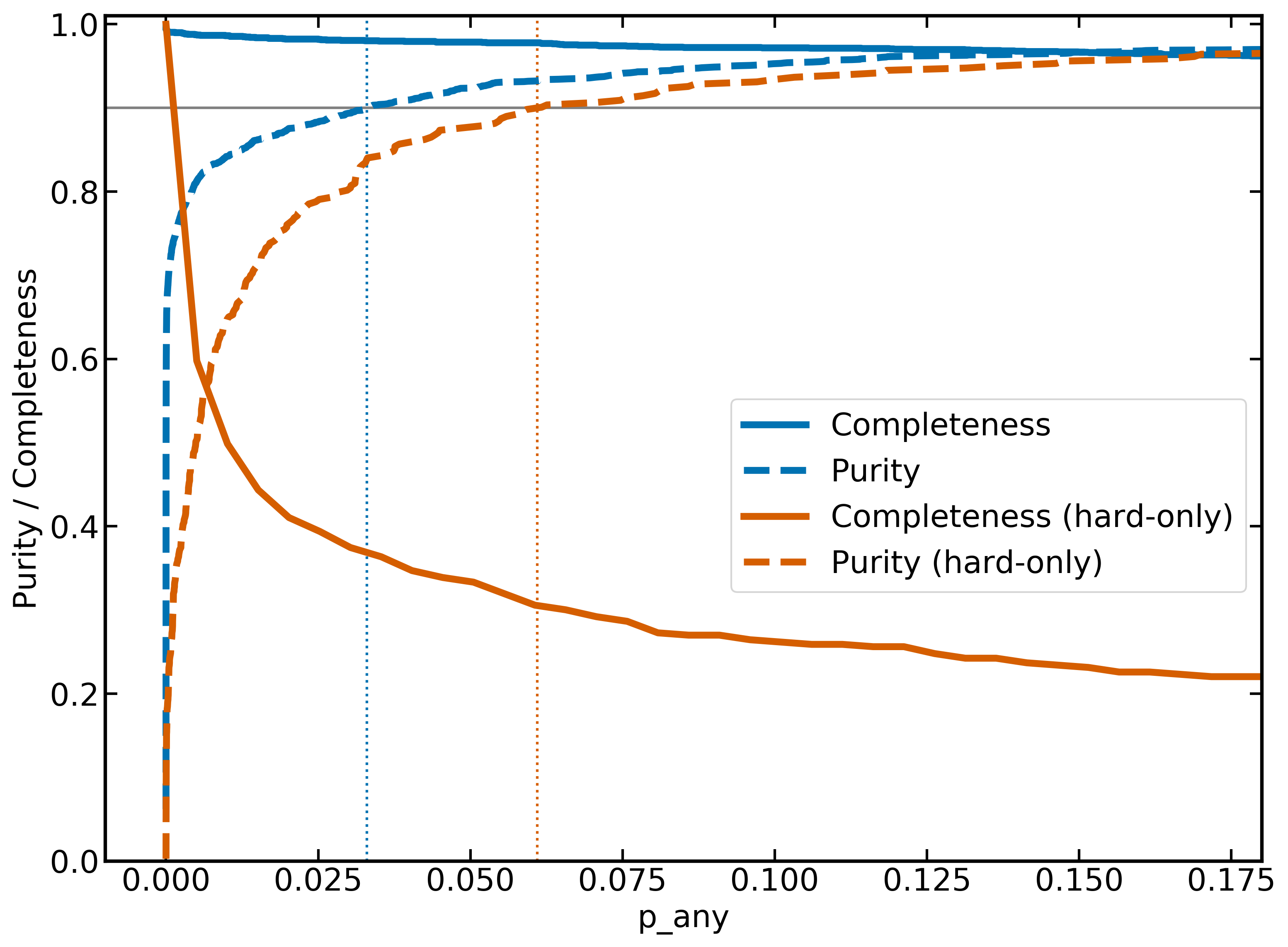}
    \caption{Purity and completeness as a function of \texttt{p\_any} values, defined as in \citet{2022Salvato}. Purity and completeness for the sample with soft X-ray detections are shown as blue solid and dashed lines, respectively, and purity and completeness for the hard-only sample are shown as orange solid and dashed lines, respectively. The horizontal grey line shows a purity of 0.9, and vertical dotted lines in corresponding colours indicate the p\_any values for the 90\% purity cuts. }
    \label{fig:pany_softhard}
\end{figure}

To define a desirable \texttt{p\_any} threshold to use to define a secure counterpart, purity and completeness curves are constructed. The completeness is defined as the fraction of sources which have a real \texttt{p\_any} greater than a certain threshold, whereas purity is defined as the fraction of spurious sources which have \texttt{p\_any} greater than a given threshold. These are shown in Fig.~\ref{fig:pany_softhard}, with, the hard-only sources shown in orange, and the sources which also have soft detections are shown in blue. Completeness curves are shown as solid lines, and purity curves are shown with dashed lines. To achieve a purity of 90\%, a threshold of \texttt{p\_any} $>0.033$ should be used for the hard sample, and \texttt{p\_any} $>0.061$ for the hard-only sample, indicated with horizontal lines in corresponding colours. The hard sample has a completeness of $\sim98\%$ at this purity, and the hard-only sample has a much lower completeness of $\sim30\%$. This is consistent with the much higher expected spurious fraction of the hard-only sample. 

\begin{figure}
    \centering
    \includegraphics[width=0.95\columnwidth]{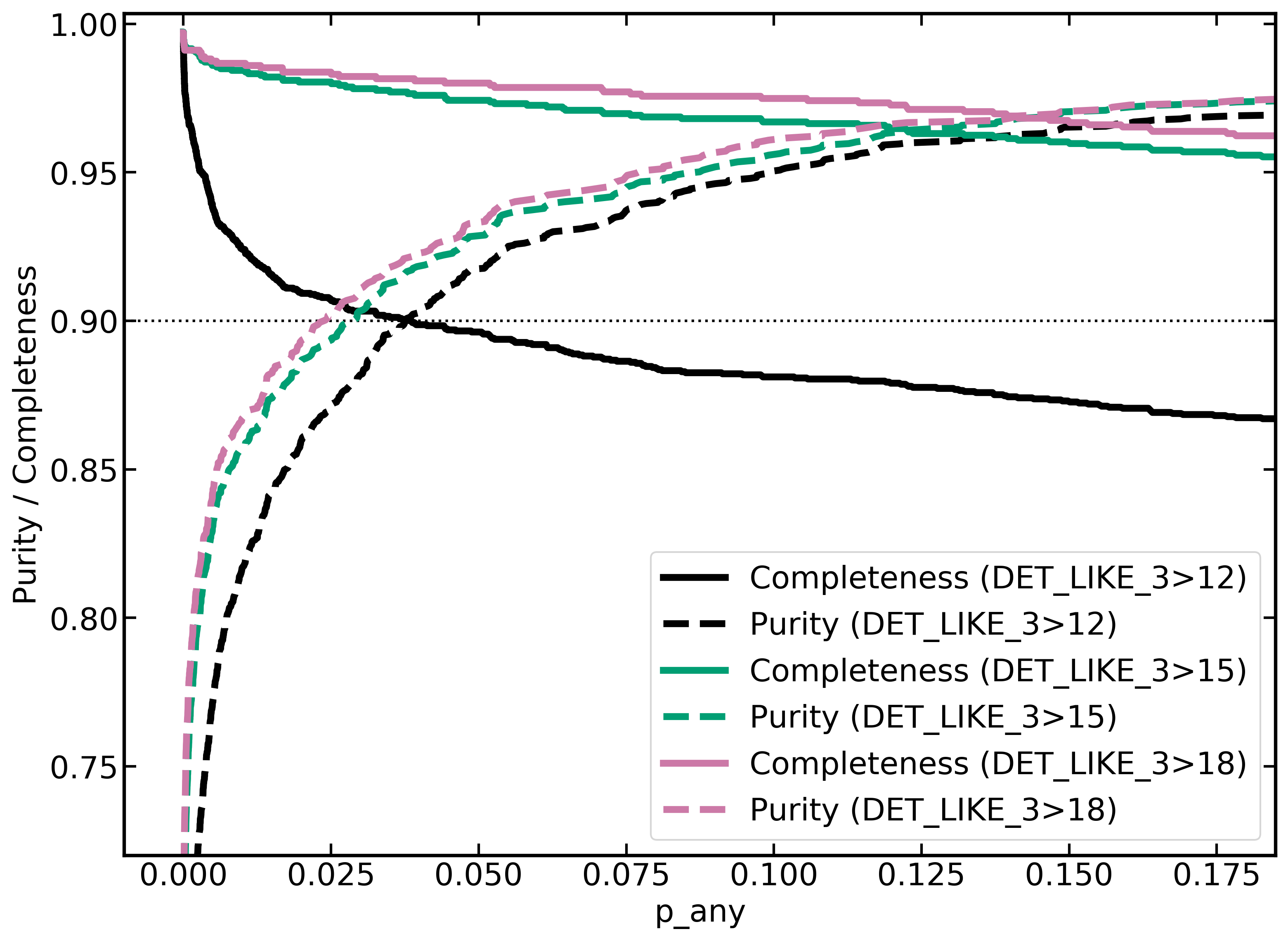}
    \caption{Purity (dashed lines) and completeness (solid lines) as a function of \texttt{p\_any} values, defined as in \citet{2022Salvato}. The black, green and pink lines show sub-samples of sources with \texttt{DET\_LIKE\_3}>12, \texttt{DET\_LIKE\_3}>15 and \texttt{DET\_LIKE\_3}>18, respectively.}
    \label{fig:pany_det}
\end{figure}

The dependence on the counterpart quality as a function of detection likelihood (\texttt{DET\_LIKE\_3}) is shown in Fig.~\ref{fig:pany_det}. Three different thresholds are shown: for \texttt{DET\_LIKE\_3}>12, 10\% of X-ray detected sources are expected to be spurious; for \texttt{DET\_LIKE\_3}>15, 2.5\% of X-ray detected sources are expected to be spurious, and for \texttt{DET\_LIKE\_3}>18, $<<1\%$ of X-ray detected sources are expected to be spurious. The purity and threshold curves are much higher when these higher \texttt{DET\_LIKE\_3} cuts are applied. These cuts also effectively remove all the hard-only sources, which typically have lower \texttt{DET\_LIKE\_3} (e.g. Fig.~\ref{fig:d3}). A highly pure, complete and very securely associated sample could be selected using higher \texttt{DET\_LIKE\_3} and \texttt{p\_any} cuts.

\begin{figure}
    \centering
    \includegraphics[width=0.95\columnwidth]{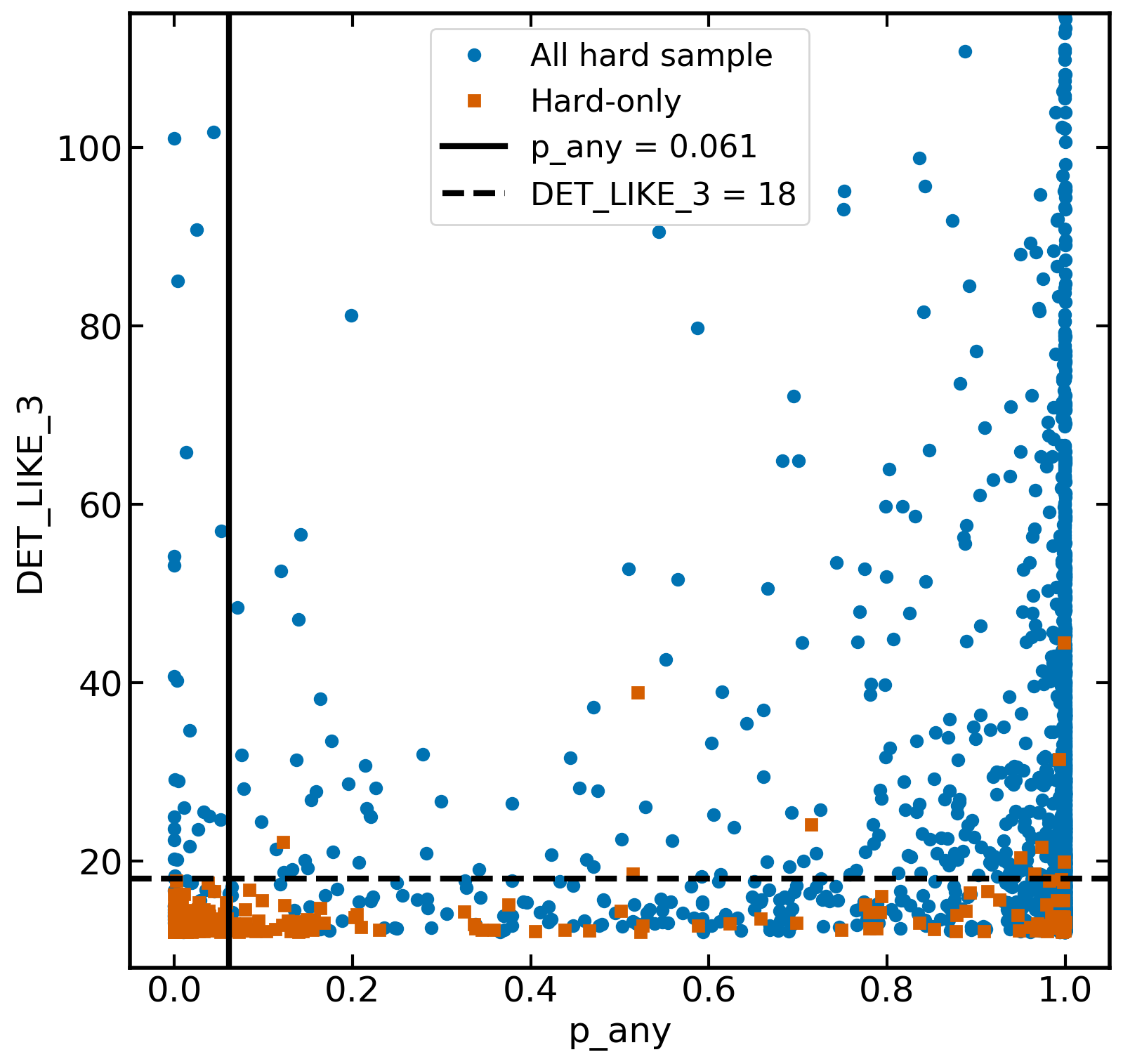}
    \caption{Distribution of \texttt{DET\_LIKE\_3} values as a function of \texttt{p\_any}, shown for both the hard-only sample (orange) and the remainder of the hard sample (blue). The vertical black line indicates the 90\% purity cut on p\_any for the hard-only sample, and the horizontal dashed line shows a high-fidelity cut of \texttt{DET\_LIKE\_3}>18. }
    \label{fig:pany_hard}
\end{figure}

The \texttt{p\_any} values with respect to the \texttt{DET\_LIKE\_3} values for the hard-only (orange squares) sources are shown in Fig.~\ref{fig:pany_hard}, alongside the full sample of sources with counterparts. A \texttt{DET\_LIKE\_3} of 18 is indicated with a black solid line, above which X-ray detections have a very low (much less than 1\%) chance of being spurious. A \texttt{p\_any} of 0.061 is also shown with a black dashed line, which indicates the 90\% purity cut on \texttt{p\_any} for the hard-only sources. Predictably, many of the low \texttt{DET\_LIKE\_3} sources also have low \texttt{p\_any}, probably because many are spurious. Most sources with \texttt{DET\_LIKE\_3} larger than 18 have secure counterparts. Of those which don't, a majority are located in the SEP, where both X-ray and LS10 counterparts are problematic and likely the results should not be trusted.  

Applying these findings, we define sources with a good counterpart as having \texttt{p\_any}>0.033, and hard-only sources with good counterparts as having \texttt{p\_any}>0.061, selecting a total of 2547 sources (of which 111 are hard-only sources) with good counterparts. These definitions are used in the following sections. 

\begin{figure}
    \centering
    \includegraphics[width=0.95\columnwidth]{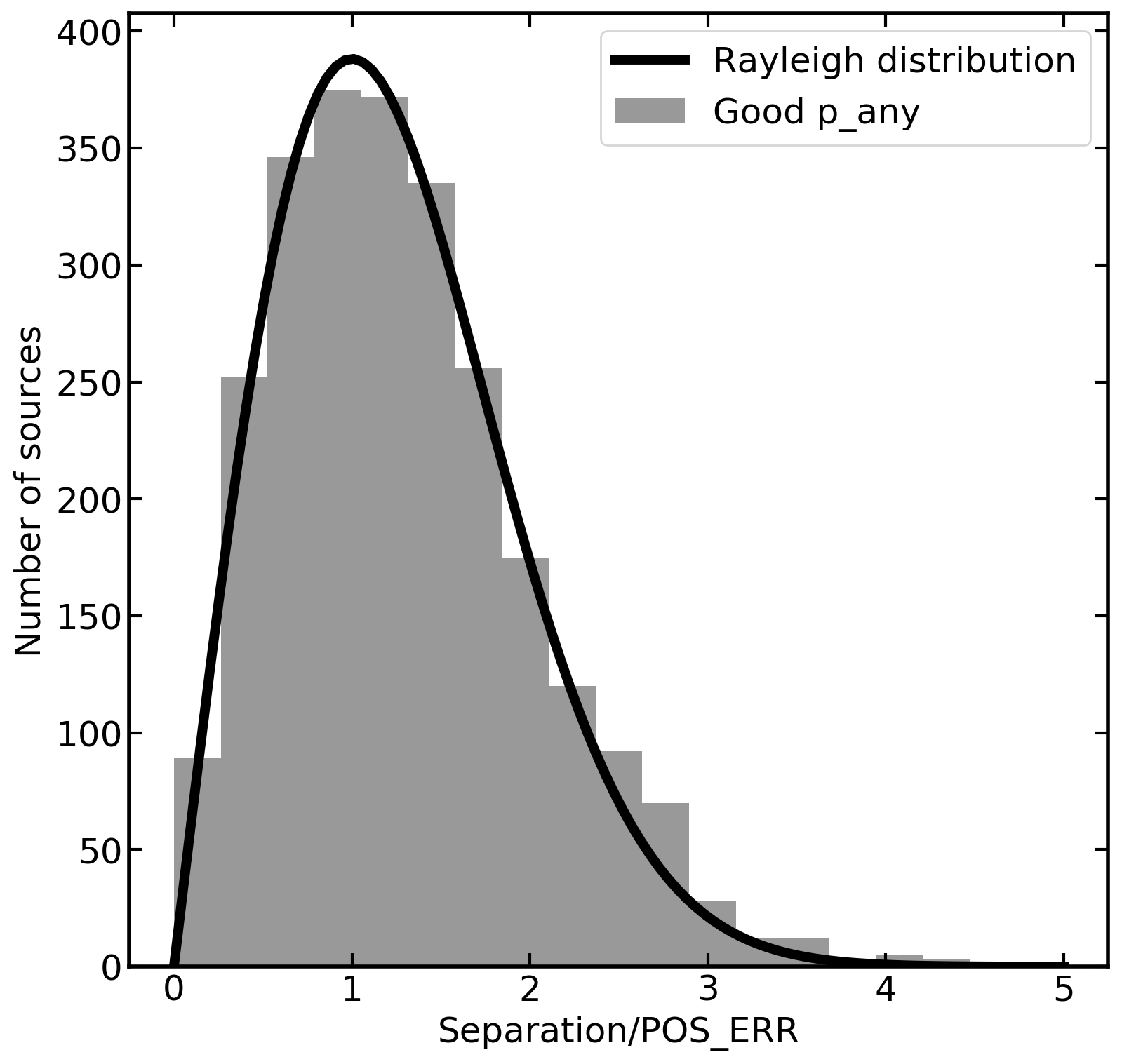}
    \caption{Distribution of separations between the optical and X-ray positions, divided by the X-ray positional errors. The sources with good p\_any are shown in grey, and the expected Rayleigh distribution is shown with a solid black curve. }
    \label{fig:sep}
\end{figure}

If the errors on the X-ray coordinates and the matching algorithm are accurate, then the the angular separations of the LS10 counterparts from their eROSITA X-ray counterparts, normalised by the position errors, is expected to be modelled by a Rayleigh distribution. This is tested in Fig.~\ref{fig:sep}, and indeed, the true distribution is in close agreement with the Rayleigh distribution.

\subsection{Spectroscopic redshifts}
Following the method of \citet{2020Ider}, a compilation of spectroscopic redshifts was constructed from a large variety of literature sources. The specific paper queries can be found in the catalog released with this work. Specific catalogs, queries and selection criteria are further described in Appendix D of Kluge \et (submitted).

The catalog includes the RA and Dec taken from the Legacy Survey sources, which are matched within one arcsecond to the counterpart LS10 positions. The separations are typically less than 0.3 arcseconds, suggesting that the spec-z are indeed being reported for the LS10 sources associated to the X-ray counterparts. In total, 715 sources with good \texttt{p\_any} have a spec-z which is available in this catalog, 710 of which have $z>0.002$ and can be considered extragalactic sources. Then, to further increase the number of spec-z available, we also match our optical counterparts to the spec-z in NED using a one arcsecond match radius. This match yields an additional 328 NED sources, 321 with extragalactic spec-z values of $z>0.002$. 

\begin{figure}
    \centering
    \includegraphics[width=0.95\columnwidth]{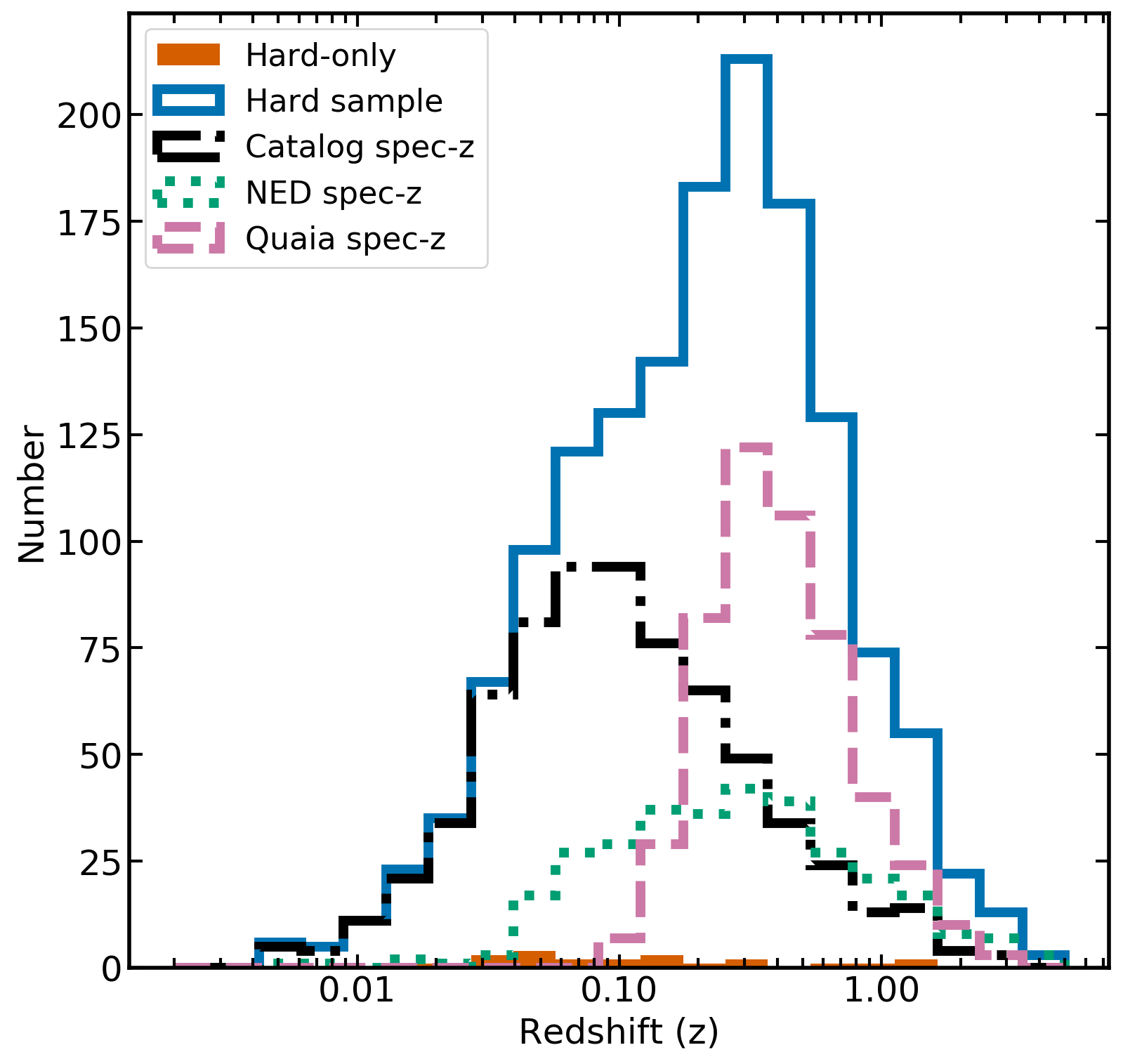}
    \caption{Distributions of spectroscopic redshifts for the eROSITA hard sample. The hard sample is shown as a blue solid line, and the hard-only sample is shown with an orange filled histogram. Spec-z from the spectroscopic redshift catalog are shown as black dash-dot lines, and NED spec-z are shown as green dotted line. The Quaia redshifts are shown with pink dashed lines.}
    \label{fig:zhist}
\end{figure}

In addition to these spec-z values, we take advantage of the inclusion of the Gaia DR3 information in the LS10 catalog to also include spec-z values from Quaia \citep{2023Storey}, the publicly available Gaia–unWISE Quasar Catalog. In this work, spec-z are measured using the low-resolution Gaia spectrometer, and are improved using WISE selection criteria. The final sample has magnitudes $G>20.5$ and shows good agreement with SDSS spec-z measurements \citep{2023Storey}. Matching directly by Gaia DR3 ID and considering sources which do not have a spec-z from the compilation or NED, 505 sources with good \texttt{p\_any} have a spec-z, all of which have $z>0.002$ are can be considered extragalactic sources, as expected from a quasar catalog. In total, this gives 1536 sources, including 23 hard-only sources, with a good counterpart and a spec-z of $z>0.002$. The redshift distribution from each source, as well as the total distribution, is shown in Fig.~\ref{fig:zhist}. 

Additionally, the agreement between the redshifts from the three catalogues used has been verified by computing the outlier fraction, where an extreme outlier is defined as:

\begin{equation}
    |z_2 - z_1| / (1+z_1) > 0.15
\end{equation}

\noindent where $z_1$ is the better-constrained redshift and $z_2$ is the lesser-constrained value. Comparing NED to the catalog, the outlier fraction is very small, at $1.5\%$. Comparing the spectroscopic redshift catalog with Quaia, the outlier fraction is higher at $\sim17\%$. This is likely related to the low resolution of the Gaia spectrograph, and the apparent lack of sources in their low redshift ($z<0.1$) SDSS comparison sample. Some of these extreme outliers are removed when beamed sources removed (see description in Section 3.6, decreasing the outlier fraction to $13\%$). A majority of the outliers are at $z>0.3$, so over interpretation of the higher-z sources is cautioned. 

\subsection{Source classification}
Learning from eFEDS, a similar source classification methodology is followed to \citet{2022Salvato}. First, all sources which have a spectroscopic redshift of $z>0.002$ are classified as extragalactic (1536 sources), while those with a spec-z of $z<0.002$ are classified as Galactic. Additionally, sources which have a parallax which is significant at the $>5\sigma$ level are classified as stars, selecting a total of 325 unique Galactic sources. 

Where no spectroscopic redshift is available, or the parallax is not significant, the optical nature of the source is examined to see if it is extended. This is done by looking at the TYPE column in the LS10 catalog. If the type is not equal to PSF, then the source is believed to be optically extended, and is thus likely to be a (nearby) galaxy. This allows us to classify 660 sources as likely extragalactic. Indeed, $\sim50\%$ of the spectroscopic extragalactic sample is optically extended, while none of the Galactic sources are, further justifying this classification.

\begin{figure*}
    \centering
    \includegraphics[width=0.93\columnwidth]{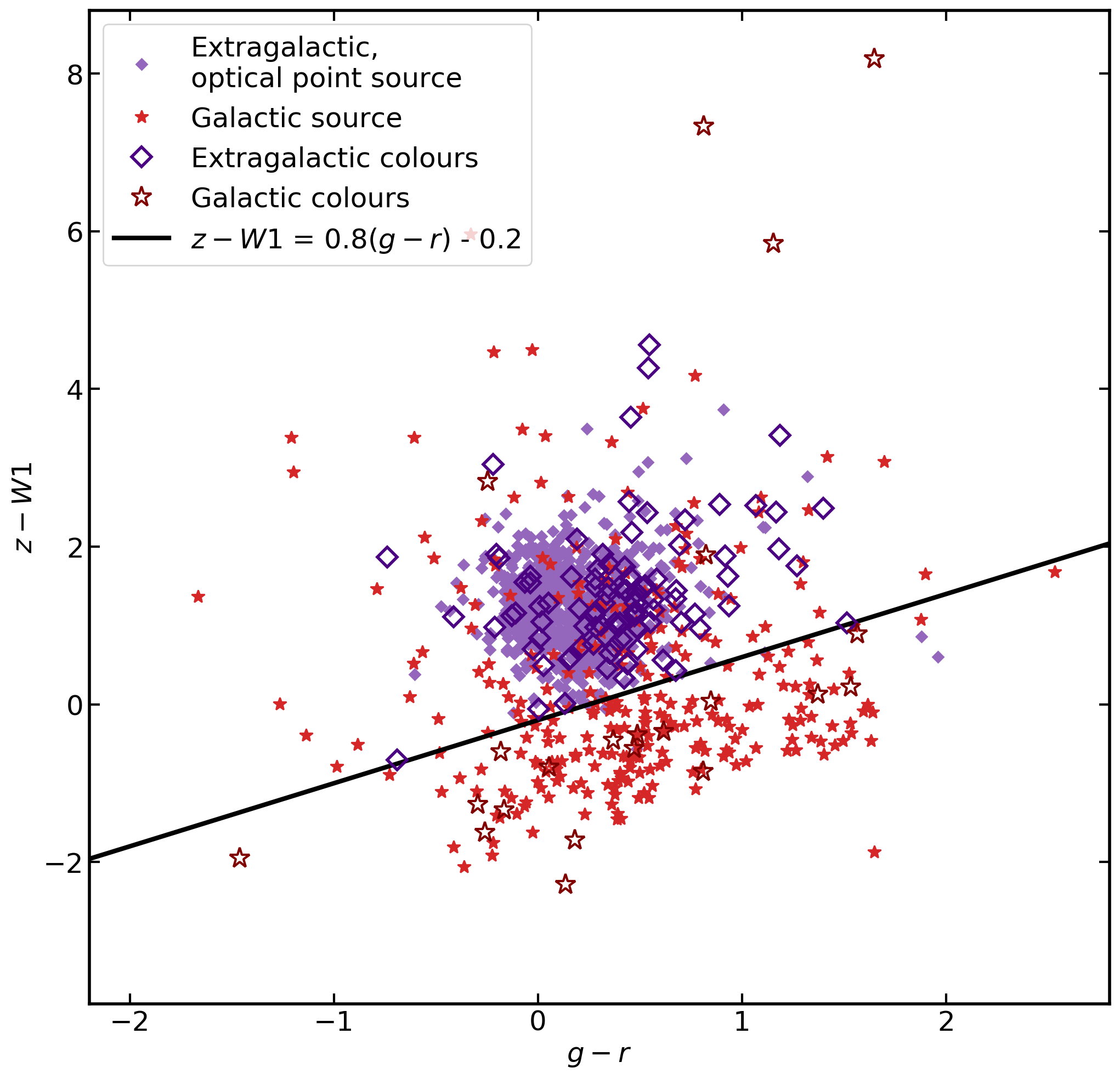}
    \hspace{2pt}
    \includegraphics[width=0.93\columnwidth]{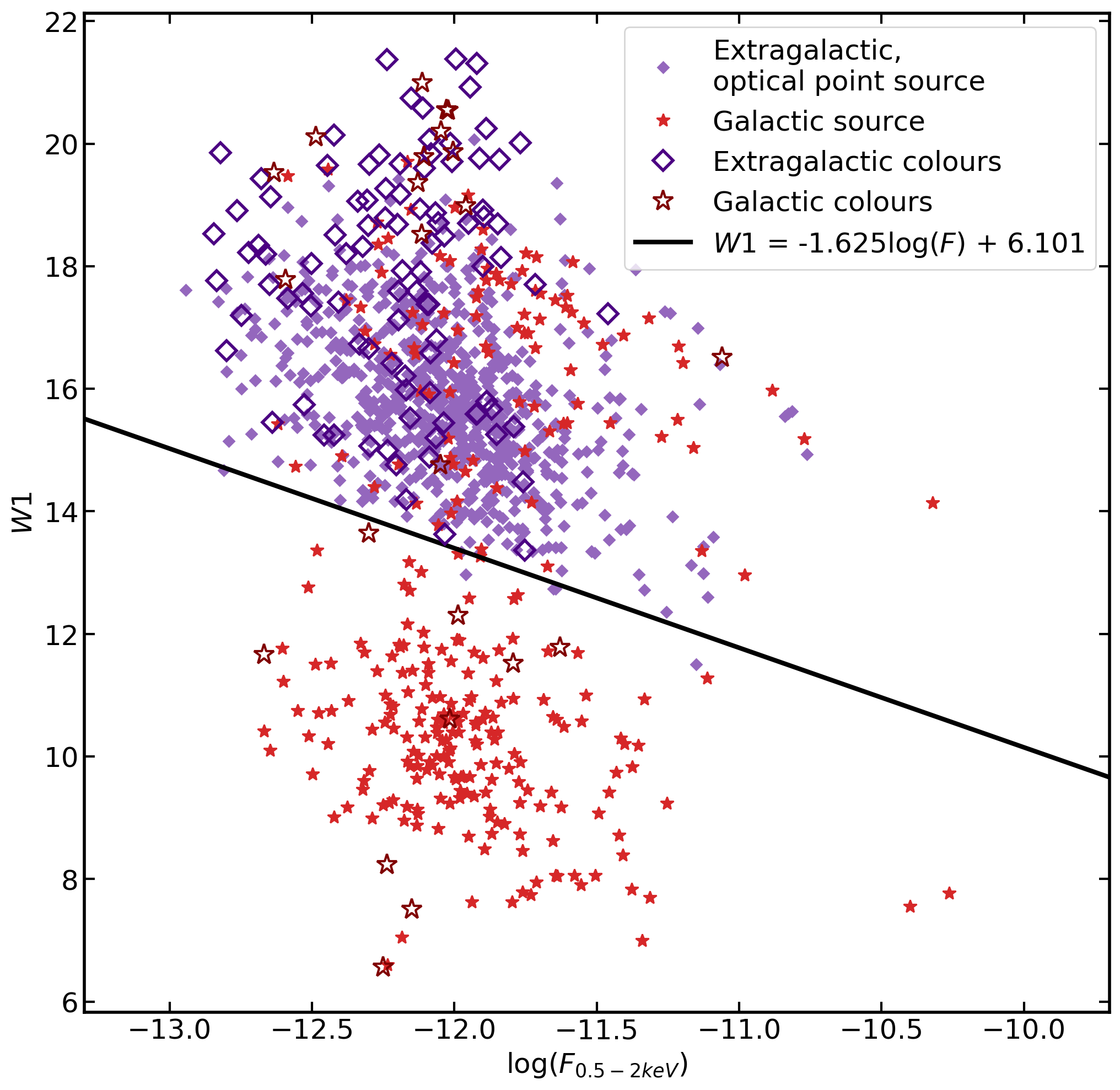}
    \includegraphics[width=0.93\columnwidth]{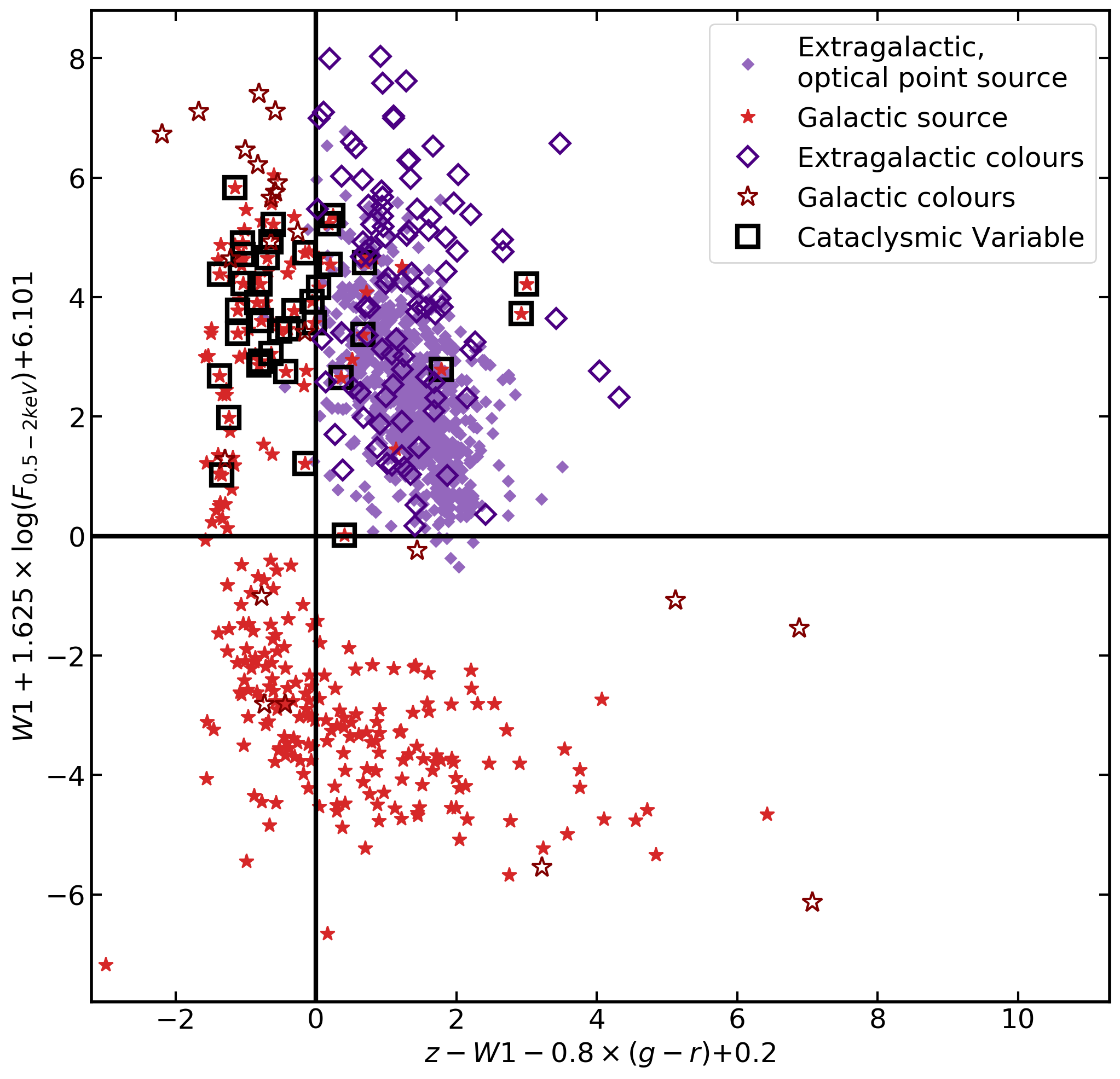}
    \includegraphics[width=0.95\columnwidth]{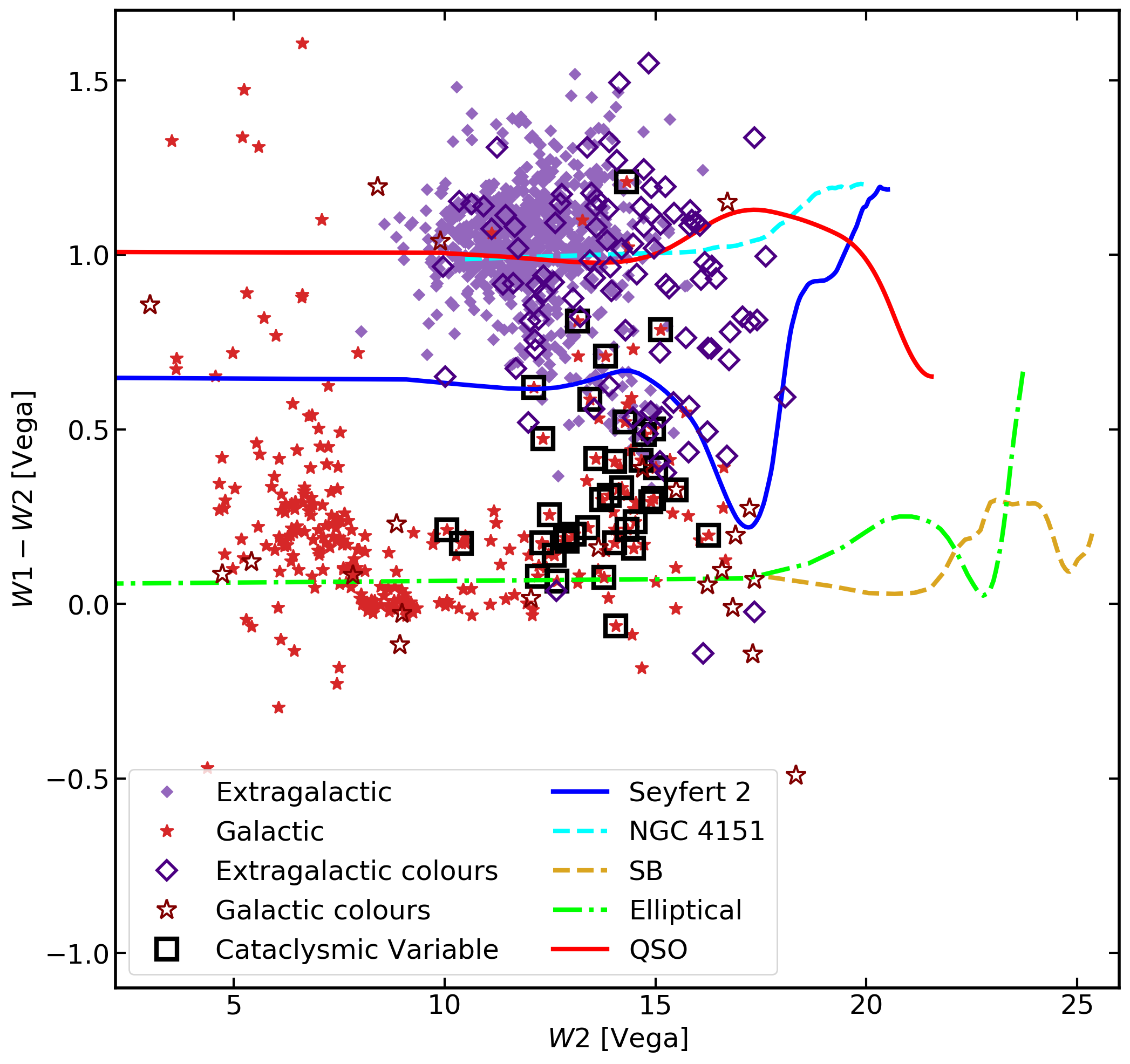}
    \caption{Colour-colour classification plots adapted from \citet{2022Salvato}. Only sources with type PSF in LS10 and sources where the signal-to-noise of $>3$ in all photometry bands are plotted. Sources classified as extragalactic (spec-z>0.002) are shown as filled purple diamonds, sources classified as Galactic (significant parallax at $>5\sigma$ or spec-z<0.002) are shown as filled red stars, sources classified as extragalatic based on their colours are plotted as unfilled, dark purple diamonds, and sources classified as Galactic based on their colours are plotted as unfilled, dark red stars. Cataclysmic variable (CV) stars are plotted with unfilled black squares on the bottom plots. The tracks on the bottom-right show the evolution of the colour-magnitude space occupied by different sources, with increasing redshift from left to right. }
    \label{fig:colours}
\end{figure*}

For remaining sources (339) without a spec-z, significant parallax, and which are not flagged as optically extended, we apply similar colour-colour cuts to \citet{2022Salvato}. The relevant colour-colour plots are shown in Fig.~\ref{fig:colours}, with different sub-samples indicated with different colours and symbols, and equations used for cuts are listed in the legend. Only sources with type PSF in LS10 and sources where the signal-to-noise of $>3$ in all photometry bands are plotted. 

The top-left panel of Fig.~\ref{fig:colours} shows the $z-W1$ and $g-r$ colours for the hard sample. The line of best-fit has the same slope as prescribed by \citet{2022Salvato}, but with an intercept of -0.2, rather than -1.2 reported in that work. This is because \citet{2022Salvato} included many local galaxies and clusters in this diagnostic, which are removed in this work by selecting only point-like sources in the X-ray and optical. This new selection is very pure, with $99\%$ of spectroscopic extragalactic sources lying in the upper region of the plot. The $W1$ vs. soft X-ray diagnostic from \citet{2022Salvato} shown in the right-hand corner of Fig.~\ref{fig:colours} are similarly pure. Here, the soft X-ray flux is computed using the hard X-ray flux which is available even for the hard-only sources, extrapolated using an absorbed power law with a photon index of 1.9 \citep[e.g.][Nandra \et submitted]{2022Liu,2023Waddell} and a column density of $4\times10^{20}$\pscm.

The bottom-left corner of Fig.~\ref{fig:colours} combines the two top figures to create the final source classification. Sources which have both colours $>0$ are classified as having extragalactic colours and plotted as unfilled, dark purple diamonds, and those with either colour $<0$ are classified as having Galactic colours and are plotted as unfilled, dark red stars. In the same plot we also mark sources classified as Cataclysmic variable (CV) stars in SIMBAD (unfilled black squares). It can be seen that many stars which lie in the extragalactic colours region are in fact CV's, where the X-ray emission may arise from the accretion process onto the white dwarf.

Summarizing, to classify the hard band selected sources with optical counterparts, we first remove 220 sources where one or both colours could not be computed, mostly due to issues with the W1 photometry, especially in the SEP region, or where sources do not have sufficient signal-to-noise in these photometric bands. Applying the colour cuts leaves 96 sources which are classified as likely extragalactic, and 23 classified as likely Galactic. Using all classification methods 2340 sources are classified as likely extragalactic or extragalactic, and 362 are classified as likely Galactic or Galactic. Out of the 2547 sources with good \texttt{p\_any}, 2114 are classified as likely extragalactic or extragalactic, and 337 are classified as likely Galactic or Galactic, indicating that $\sim80\%$ of the sample is extragalactic.

We note that the classifications based only on colour-colour classifications should be interpreted with some caution, particularly for the stars, as some are located in a region of the parameter space (upper-left quadrant of Fig.~\ref{fig:colours}) which is occupied by extended (and thus likely extragalactic) sources with Galactic-source-like colours. Furthermore, we again stress that sources with low \texttt{p\_any} may have different astrophysical origins for the X-ray and W1 photons, making the W1-X-ray colour unreliable.

Finally, the bottom-right panel of Fig.~\ref{fig:colours} shows the $W1-W2$ colour versus the W2 magnitude for these sources. Overplotted here are tracks corresponding to the colours of typical sources as they increase in redshift, akin to those presented in \citet{2022Salvato}. A majority of extragalactic sources are shown to lie in the vicinity of the QSO and NGC 4151 tracks, with very few galactic sources matching Elliptical and SB tracks. Also of interest is a smaller number of sources which appear to lie along the Seyfert 2 track, which are commonly associated with obscured AGN. These sources will be further discussed in Section 3.7. 

\subsection{Beamed AGN}
Having defined a sample of secure and likely extragalactic sources, it is also of interest to identify which of these may be associated with blazars. The X-ray emission from these sources is likely to originate in the jet rather than in the X-ray corona, and may be relativistically boosted. This means that the total spectral shape and luminosity is unlikely to be representative of the corona, so removing these from an AGN sample will improve our understanding of the true coronal emission.

To identify candidate blazars in our sample, we perform a cross-match with three catalogs/databases; CRATES, BZCAT and SIMBAD. CRATES \citep{2007Healey} is an 8.4GHz selected flat-spectrum radio quasar (FSRQ) catalog constructed using multiple radio surveys and additional follow-up work to achieve nearly uniform sky coverage of latitudes $|b|>10^{\rm{o}}$. The Roma-BZCAT catalog is a multi-frequency blazar catalog, containing both FSRQ sources and BL Lac. The 5th edition was used here \citep{2014Massaro}, containing 3561 sources, all of which have radio detections. Finally, to identify other blazar candidates, we cross-match with SIMBAD. Sources which have a type of Blazar, BLLac, Blazar\_Candidate, or BLLac\_Candidate are flagged. The information is stored in the column class\_beamed, where sources which appear in BZCAT have a class of 1, sources appearing in CRATES have a class 2, sources in SIMBAD have a class 3, and other sources have class 0. In total, 319 sources are flagged as beamed. These sources are thus removed from the AGN sample, and will be presented in Haemmerich \et (in prep). 

\subsection{Multiwavelength properties of the hard-only sources}
Of particular interest throughout this work is the hard-only sample, which is likely to consist largely of obscured AGN which lack soft X-ray emission. However, it should also be noted that the hard-only sources may have properties which are not consistent with the remainder of the main sample. In particular, it is clear that many more hard-only sources are in fact spurious, and that even some sources which have a good counterpart may still be chance associations of galaxies and spurious hard X-ray detections. 

\begin{figure}
    \centering
    \includegraphics[width=0.95\columnwidth]{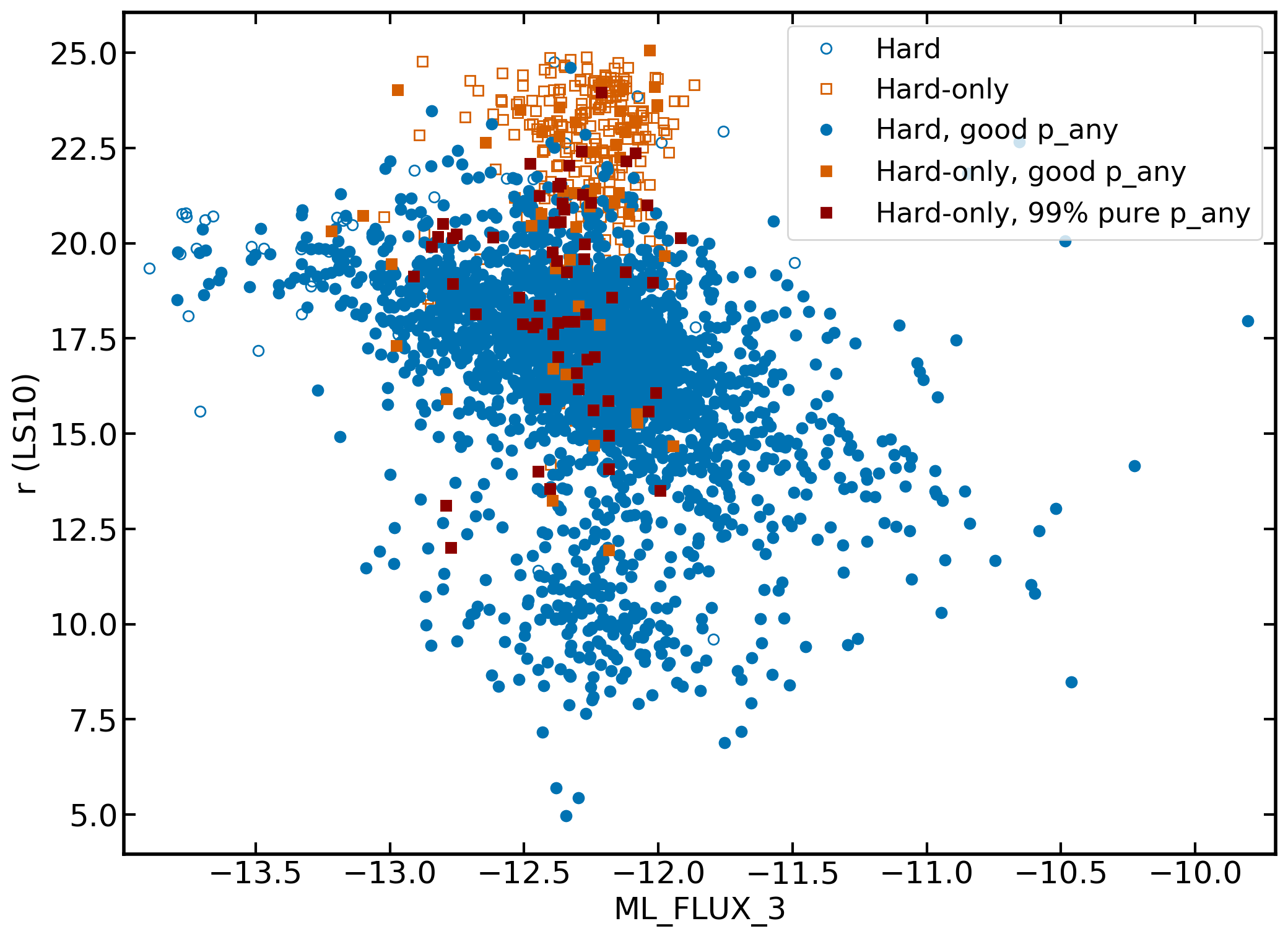}
    \caption{De-reddened r magnitudes and hard X-ray ($2.3-5\kev$) fluxes for the hard sample. All hard sources with \texttt{p\_any} less than the threshold of 0.033 are shown as blue unfilled circles, and sources with good \texttt{p\_any} are shown with blue filled circles. Hard-only sources with \texttt{p\_any} less than the threshold of 0.061 are shown as orange unfilled squares, and hard-only sources with good \texttt{p\_any} are shown with orange filled squares. A 99\% purity cut on \texttt{p\_any} at 0.61 is also applied, and these sources are shown as dark red filled squares.}
    \label{fig:rx}
\end{figure}

One visualisation of this can be seen in Fig.~\ref{fig:rx}, which shows the distribution of de-reddened r magnitude and hard X-ray fluxes for the sample. The population of stars is evident, with $R<12.5$ for typical X-ray fluxes. A very large population of hard-only sources lie in an extreme region of the parameter space, with relatively typical hard X-ray fluxes but extremely faint r band magnitudes. However, it can also be seen that many of these sources have low \texttt{p\_any}. To further investigate this, we also plot a second sub-sample with a 99\% purity cut on \texttt{p\_any}, corresponding to \texttt{p\_any} > 0.61, shown in dark red. Here it is clear that while some hard-only sources are fainter in r, other sources with $R>22$ are likely wrong associations. Further investigation optical and X-ray flux ratios of these sources is left for future work.

\begin{figure}
    \centering
    \includegraphics[width=0.95\columnwidth]{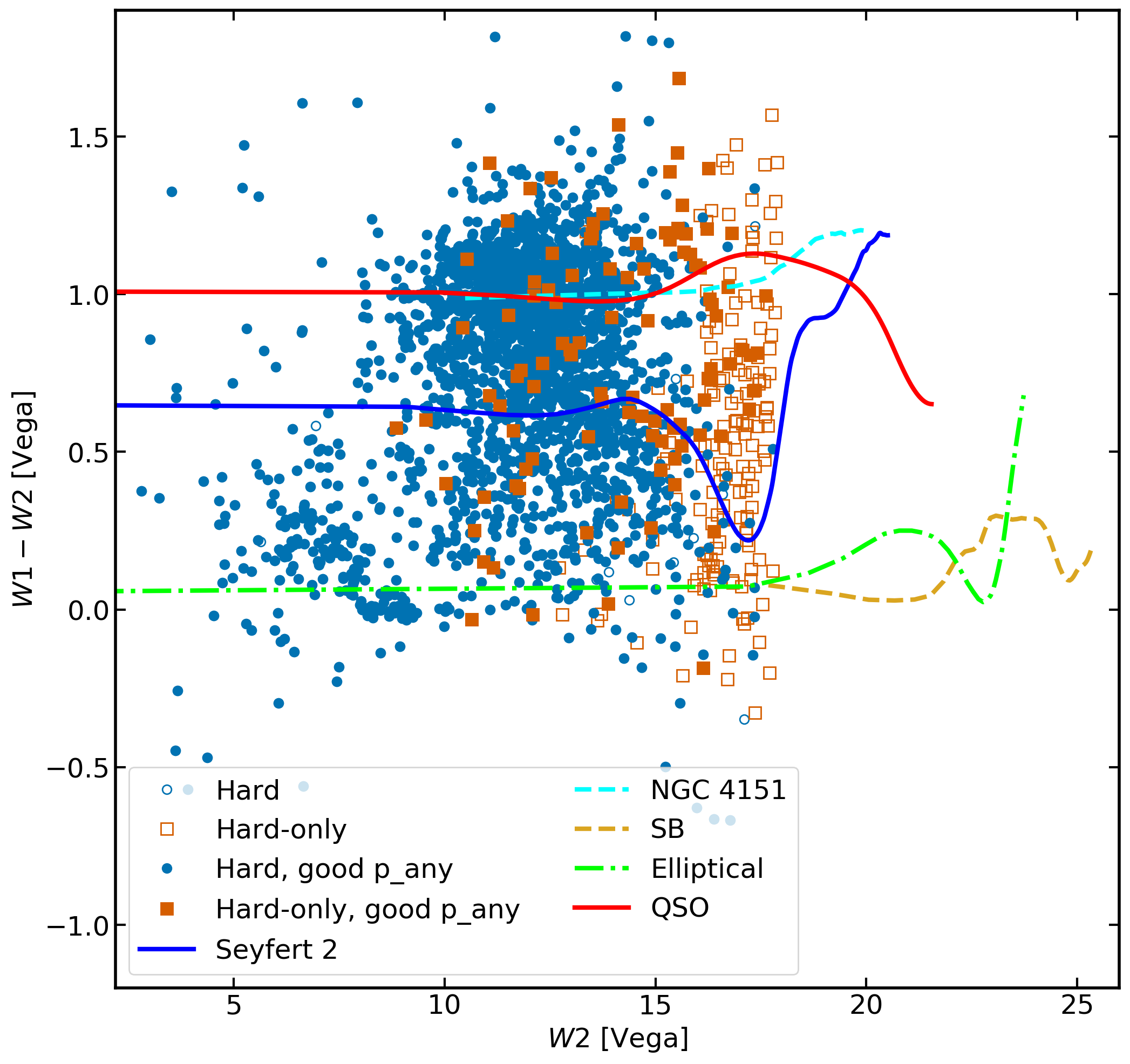}
    \caption{W1-W2 and W2 magnitudes shown for the eRASS1 hard sample. All hard sources with \texttt{p\_any} less than the threshold of 0.033 are shown as blue unfilled circles, and sources with good \texttt{p\_any} are shown with blue filled circles. Hard-only sources with \texttt{p\_any} less than the threshold of 0.061 are shown as orange unfilled squares, and hard-only sources with good \texttt{p\_any} are shown with orange filled squares. Tracks as in \citet{2022Salvato} and Fig.~\ref{fig:colours} of this work are shown in various linestyles and colours.}
    \label{fig:w1w2}
\end{figure}

Also of interest is to examine the location of the hard-only sources in the $W1-W2$ vs. $W2$ plane, and in particular in reference to the Seyfert 2 track. This is shown in Fig.~\ref{fig:w1w2}. We note that this figure includes sources which have extended PSF types, and so includes additional source as compared to the bottom-right panel of Fig.~\ref{fig:colours}. Indeed, many hard-only sources are consistent with being Seyfert 2 AGN. Some are also consistent with stellar populations, and additional sources are also in agreement with being very bright QSOs, which is worthy of further exploration. 

\begin{figure}
    \centering
    \includegraphics[width=0.95\columnwidth]{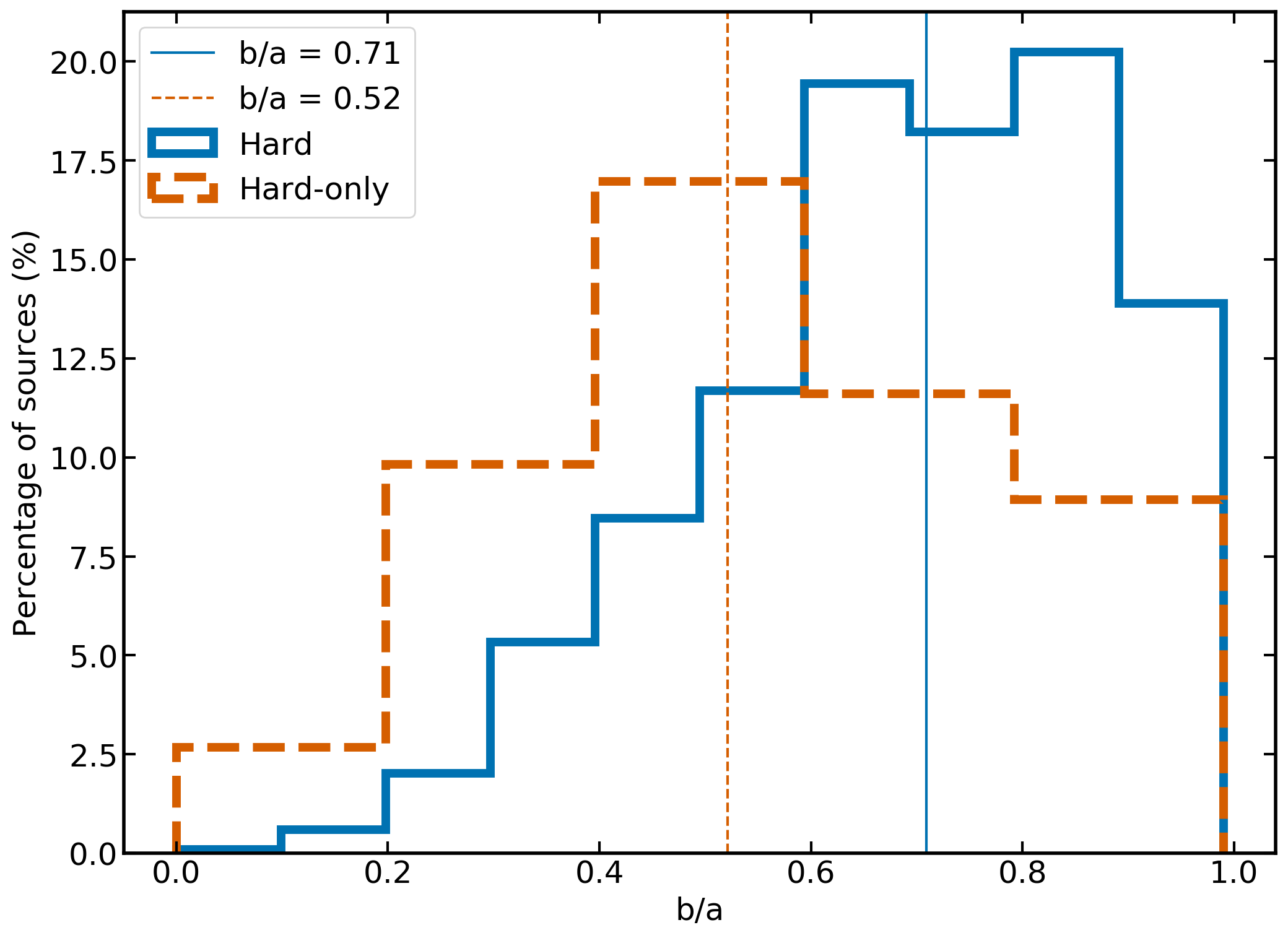}
    \caption{Ratio of the semi-major and semi-minor axes of the optical counterparts, where available. Sources are shown in blue solid lines, and sources in the hard-only sample are shown in orange dashed lines. The respective median values are indicated with vertical lines. Only sources with good \texttt{p\_any} are shown. }
    \label{fig:ba}
\end{figure}

Additionally, the optical light profiles of sources which do not have an optical type of PSF are fitted in order to measure the ellipticity of the source \citep{2019Dey}. For sources where this information is available, the ratio between the semi-major and semi-minor axes, a/b, is shown in Fig.~\ref{fig:ba}. Hard-only sources are shown separately, and are binned up in order to better compare the distribution of sources to the rest of the hard sample. The median b/a for the sources with both hard and soft detections is 0.71, whereas for hard-only sources, it is 0.52. This could imply that the hard-only sources tend to preferentially found in galaxies which are edge-on. This interpretation will be discussed further in Sections 8.2 and 9.2. 

\section{Counterparts outside the legacy survey footprint}
As an additional complement to the LS10 counterparts, it is also of interest to attempt to identify counterparts outside of the legacy survey area. A comprehensive treatment like for LS10 with other surveys goes beyond the scope of this work. However, for the hard-detected sources which also have soft emission, the positional uncertainties are very small (c.f. Fig.~\ref{fig:radecerr}). This enables a simple positional match to archival catalogs. These sources are primarily located at low Galactic latitudes which are not covered in LS10, meaning that this sample consists of AGN located behind the Galactic plane, making them of particular interest as it is more difficult to select AGN here. To do so, we first select 1710 sources which are outside of the LS10 MOC, have no flags, and are not hard-only. We then perform a positional match to public QSO catalogs; the Gaia/Quaia AGN sample \citep{2023Storey}, the Gaia/unWISE catalog \citep{2019Shu} and the Million Quasar Catalog \citep[MilliQSO;][]{2023Flesch}. In all, we recover 692 matches out of 1710 sources, such that 40\% of the sample outside LS10 has an QSO match. Out of these, 339 have spec-z from Quaia, and 349 have photo-z from the other catalogs \citep{2023Flesch,2023Storey}. More details of this procedure are described in the Appendix, and these sources are also shown in Fig.~\ref{fig:lz}. 

\section{Comparison with Swift-BAT AGN}
\subsection{Match to Swift-BAT}
The Swift Burst Array Telescope (BAT) is sensitive in the $15-150\kev$ energy range, and is thus highly complementary to many other contemporary X-ray missions. With a large field of view and rapid slew capability, the mission was designed to provide rapid follow-up for Gamma-ray bursts and other highly energetic events. The BAT instrument has also surveyed the sky in X-rays, with the first 105 months of operation providing a sample of 1632 hard X-ray detections \citep{2018Oh} with a sensitivity of $8.40\times10^{-12}$\fluxcgs\ in the $14-195\kev$ band, over 90\% of the sky. Counterparts have also been identified for these sources in \citet{2018Oh} and \citet{2017Ricci}, and sources are classified based on their optical spectra or using other multiwavelength properties. This sample is of interest for comparison with eROSITA, as they are in highly complementary energy ranges, and both surveys cover bright, nearby sources, especially AGN.

To find matches with BAT, we first perform a 60 arcsecond cross-match between the eROSITA sky positions and the Swift BAT 105 month catalog counterpart positions provided by \citet{2018Oh}. By examining the distribution of separations, we determined that a 10 arcsecond match radius was appropriate. It should be noted here that while matching the X-ray positions directly to an optical catalog is not appropriate in this sample, it is extremely unlikely to have a spatially coincident BAT and hard X-ray selected eROSITA source which do not have the same physical origin, justifying this matching procedure. Out of the 849 BAT sources in the eROSITA-DE area, we found 487 matches, of which 481 are point sources and 456 are point sources with no eROSITA flags. There are no duplicates in this matching analysis. Additionally, 250 matched sources were inside the LS10 area. For the subsequent analysis, we consider all 487 matches for comparison. 

\begin{figure}
    \centering
    \includegraphics[width=0.95\columnwidth]{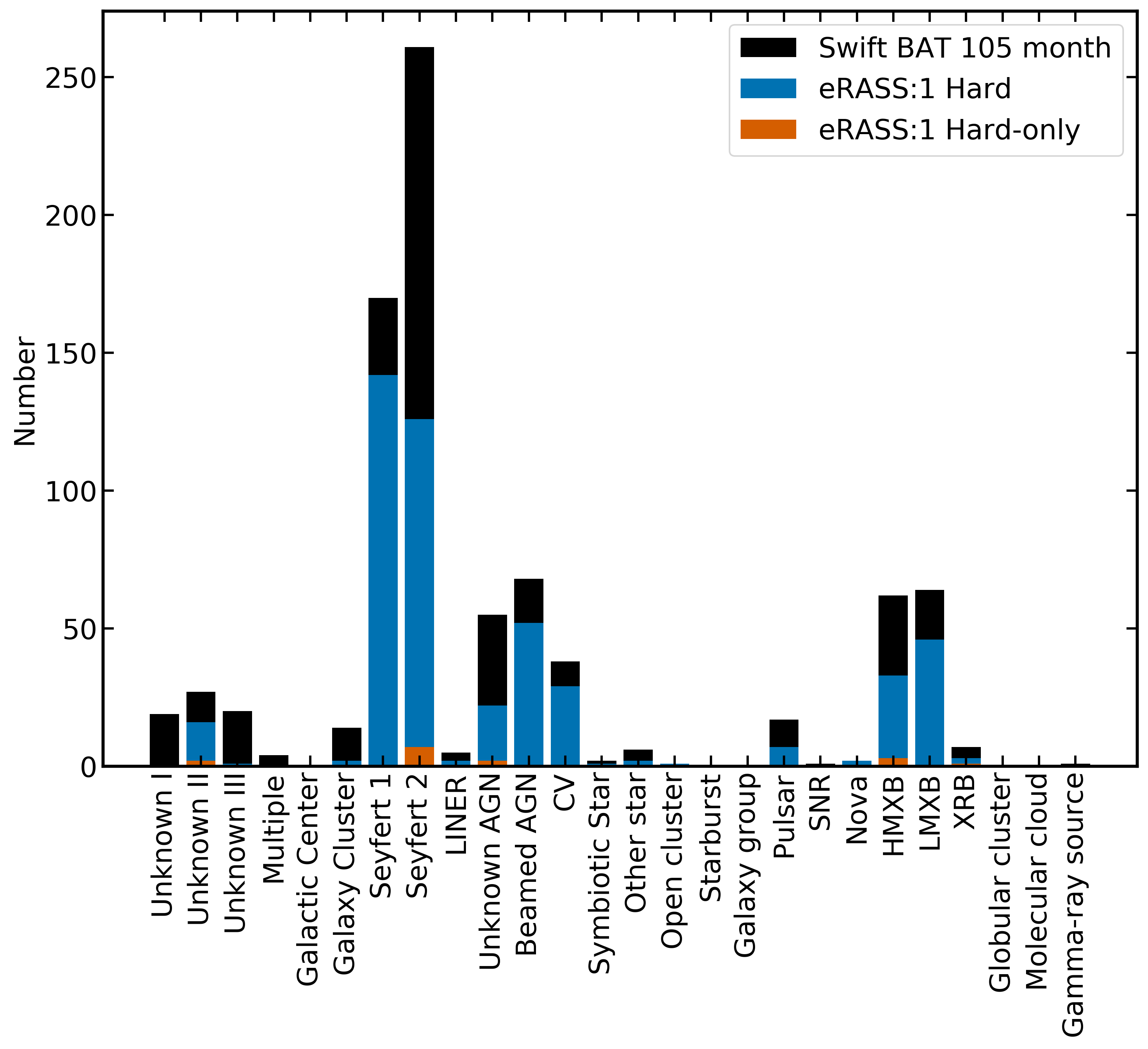}
    \includegraphics[width=0.95\columnwidth]{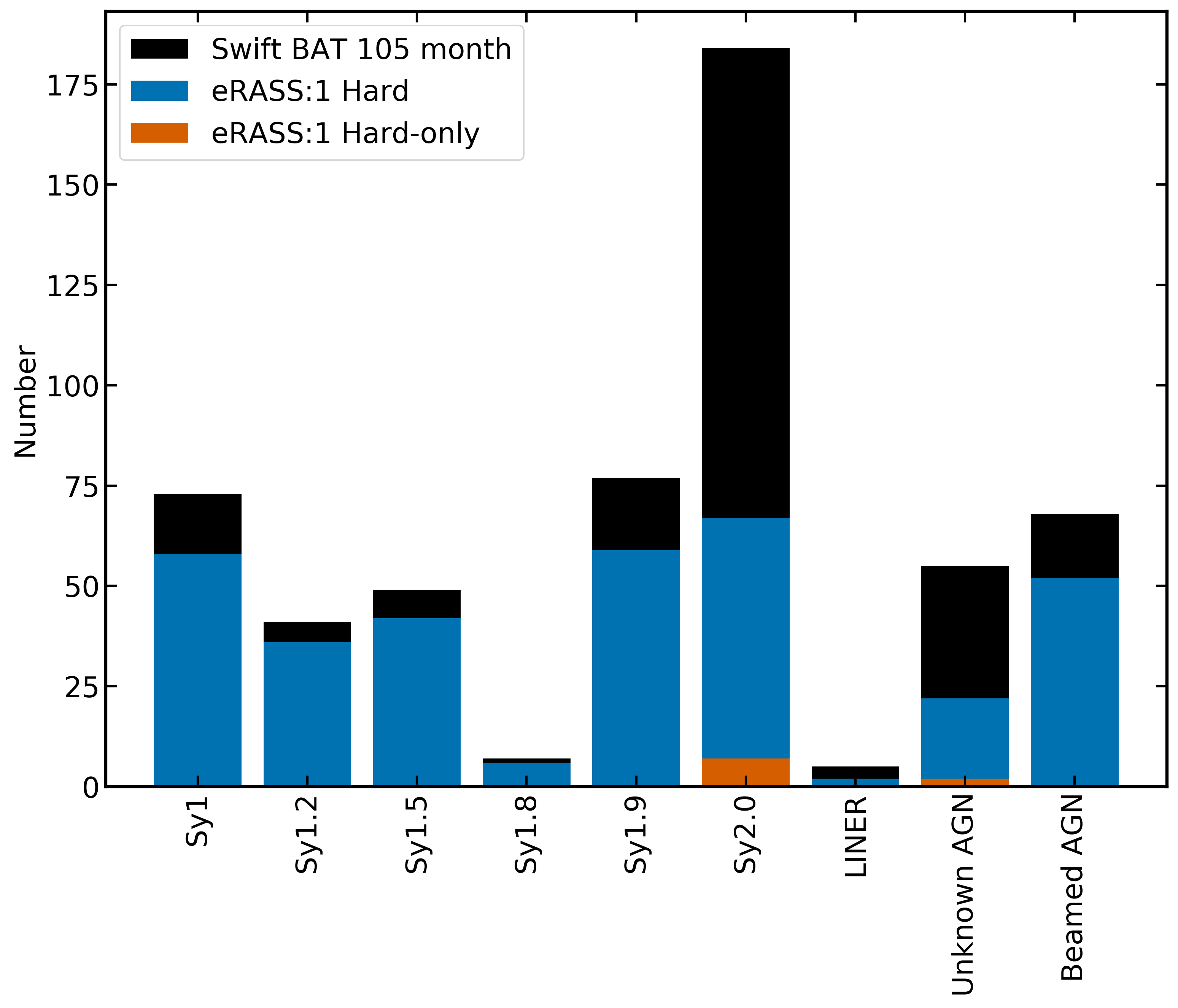}
    \caption{Top: Distribution of BAT 105 month catalog sources in the eROSITA-DE sky according to type (black), and of matched eROSITA sources according to BAT type (blue). Bottom: Distribution of BAT 105 month catalog sources according to BAT class (black), and of matched eROSITA sources according to BAT class (blue). Hard-only eROSITA sources are shown in orange. }
    \label{fig:bat_det}
\end{figure}

It is then of interest  to determine which types of BAT sources are detected by eROSITA. BAT sources are assigned a type by searching for publicly available optical soft X-ray and spectra, and the procedure is described in \citet{2018Oh}. Fig.~\ref{fig:bat_det} (top) shows the numbers of matched and non-matched sources in the eROSITA-DE sky, separated by the different source classes in the BAT 105 month catalog \citep[c.f. Table 1 of][]{2018Oh}. A large fraction of the sources detected with eROSITA are Seyfert 1 AGN with broad optical lines (142), followed by Seyfert 2 AGN (126) and beamed AGN (52). A large fraction of CVs, X-ray binaries, and unknown AGN (where the optical spectra and type are not available/known) are also detected, as well as sources of Unknown class II, which are sources with previous soft X-ray detection, but with unknown type. Of the 15 matches with the hard-only sample, seven are Seyfert 2, two are AGN of unknown class, two are unknown class II sources, and four are X-ray binaries.

Beyond type 1 and 2 Seyfert galaxies, examining the optical spectra can also allow further sub-division into the classes Seyfert 1, 1.2, 1.5, 1.8, 1.9 and 2.0, classified based on the strength of the broad optical lines. Fig.~\ref{fig:bat_det} (bottom) shows the numbers of matched and non-matched sources in the eROSITA-DE sky, according to these optical classes. It is clear that eROSITA is highly sensitive to detecting the relatively un-obscured, type 1-1.9 AGN, but is much less sensitive to detecting type-2 AGN, which is sensible as these sources are likely to be heavily obscured and thus very dim or absent in the eROSITA bandpass. The seven hard-only Seyfert 2s and two hard-only unknown AGN are also shown.

\begin{figure}
    \centering
    \includegraphics[width=0.95\columnwidth]{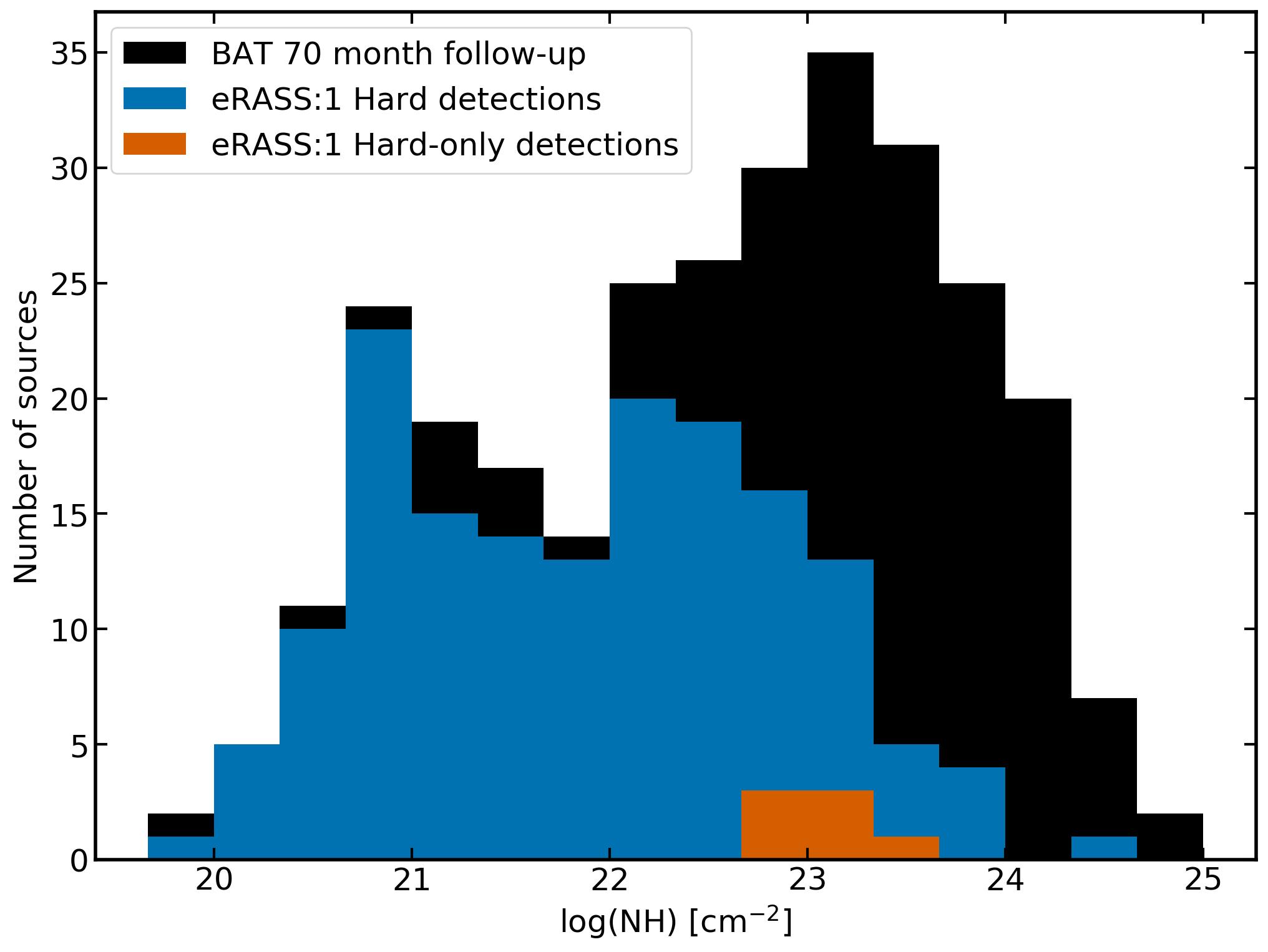}
    \caption{Distribution of column densities log(N$_{\rm H}$) of sources with soft X-ray detections in the BAT 70 month catalog, shown for sources in the eROSITA-DE sky (black) and for those matched with eROSITA sources (blue). Hard-only eROSITA sources are shown in orange. }
    \label{fig:nh}
\end{figure}

An earlier release of the BAT catalog (the BAT 70 months catalog) also reported information on the soft X-ray properties, including the column density, of BAT-detected AGN, where available \citep{2017Ricci}. We therefore match the eROSITA hard sample to this catalog, finding 273 matches out of 435 sources in the eROSITA-DE sky. This match can also be justified by assuming that spatial coincidence of an \ero\ hard source with a Swft-BAT source is extremely unlikely. For sources where the soft X-ray follow-up detection (with e.g. Swift XRT, Chandra, XMM-Newton) allows for the constraint of the column density (293 sources), we also identify matched sources in eROSITA (159). The histogram of these column densities is shown in Fig.~\ref{fig:nh}. This is not representative of the true underlying column density of AGN in the local Universe, as many sources did not have significant data quality to constrain the column density and have been removed from this plot, suggesting that unobscured sources may be underrepresented. However, the analysis shows that eROSITA is capable of detecting sources which, at the time of the observations for the Swift-BAT program, had column densities in excess of $10^{24}\pscm$, and appears to reliably detect sources up to column densities of $10^{23}$\pscm, where the detection fraction appears to fall to $\sim50\%$. Of key interest as well is the distribution of the seven hard-only sources which also appear in this catalog. All seven have column densities of $>10^{22.5}\pscm$, and a majority (4/7) are above $10^{23}\pscm$, further demonstrating that eROSITA can select obscured AGN.

The distribution also shows a bimodality in column densities with a dip at $\sim10^{22}\pscm$, which is also visible in the eROSITA matched sources and is also found in eFEDS (Nandra \et submitted). This suggests the hard X-ray selected eROSITA sample will be able to give further constraints of column densities of AGN in the local Universe. 

\subsection{Comparison to the LS10 positions}
In order to assess the validity of the LS10 counterparts, an independent check is to compare the positions to those derived from the BAT counterpart positions. To do so, we select the 250 sources which had BAT 105 month catalog matches which are in the LS10 area, and remove three sources which were flagged as problematic in the X-ray catalog. We remove one additional source, NGC 1365, which has missing photometry in the LS10 catalog and is thus not found as a counterpart. We then match these to the X-ray sources, and measure the separation between the BAT counterpart positions and the LS10 counterpart positions. Of these 246 sources, 227 have separations of less than two arcseconds, and are thus determined to be associated with the same optical source. For the remaining 19 sources, we visually inspect the LS10 image, and verify each source in SIMBAD. The maximum separation is $\sim8$ arcseconds. All sources appear to be consistent, and every difference in position seems to be related to small differences between the coordinates in the BAT catalog versus other catalogs. This could be a rounding or entry issue in the catalog. Crucially however, the analysis lends additional faith to the \texttt{NWAY} counterparts, and suggests that the training sample from Salvato \et (in prep.) is sufficient for use in this work. More discussion will be given in Section 8.1. 

\section{Comparison with the Piccinotti AGN sample}
The HEAO 1 experiment A2 \citep{1979Rothschild} performed an X-ray survey in the $2-10\kev$ band, covering 8.2sr of the sky at galactic latitudes $|b|>20^{\rm{o}}$ with a limiting sensitivity of $3.1\times10^{-11}$\fluxcgs. A total of 85 sources were detected and identified, 31 of which are AGN, forming a complete hard X-ray selected sample consisting of 27 Seyfert 1s, three Seyfert 2s, and one blazar (3C 273) \citep{1982Piccinotti}. These well-studied AGN are worthy of analysis with eROSITA, as this may be considered as the previous best, all-sky, hard ($2-10\kev$) selected sample of AGN. 

Out of the 31 AGN in the Piccinotti sample, 20 are found in the eROSITA\_DE sky. All of these sources are detected by eROSITA (both in the main and hard sample) and can be found in Table~\ref{tab:pic}. eROSITA-derived $2.3-5\kev$ fluxes and $2-10\kev$ Swift BAT follow-up fluxes are given for each source, as well as the fluxes derived in HEA0-1. It can immediately be seen that many of the eRASS1 and BAT fluxes are below the values measured in HEAO1, and indeed many are below the HEA01 limiting flux. This may be due to intrinsic variability, as HEAO1 detected only the brightest sources in the X-ray sky and likely detected only AGN in a bright state. The relative fluxes measured between missions are also model depedent, and may be in better agreement with improved spectral modelling.

\begin{figure}
    \centering
    \includegraphics[width=0.95\columnwidth]{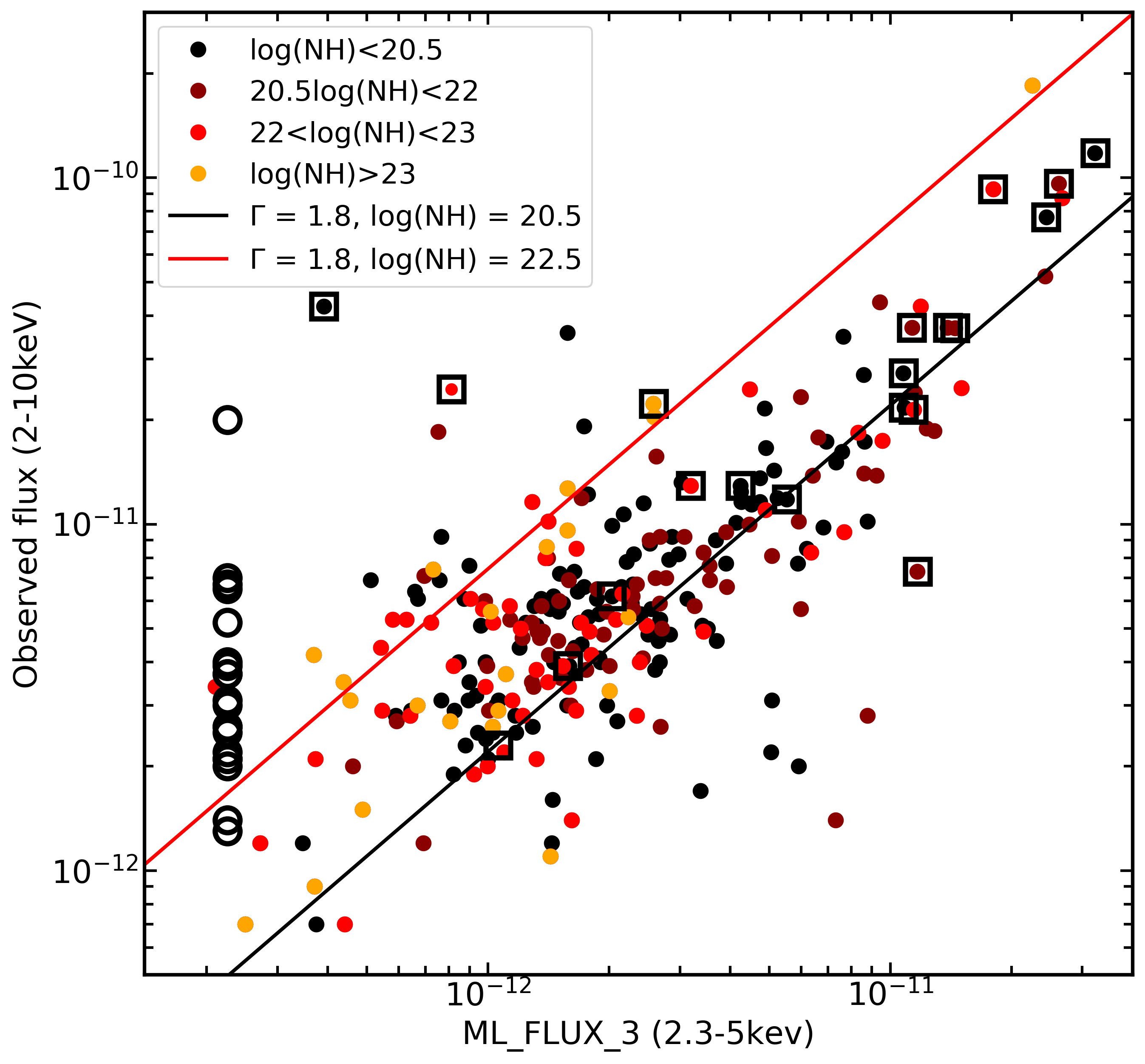}
    \caption{eROSITA $2.3-5\kev$ fluxes (ML\_FLUX\_3) and BASS fluxes ($2-10\kev$) for Piccinotti AGN (black, open squares), as well as the other sources in the BAT 70 month sample (circles). Open circles show the sources with non-detections in eROSITA which have measured column densities of N$_{\rm H}$<$10^{22.5}\pscm$. The black solid line shows the equivalent 1-to-1 line assuming a photon index of $\Gamma = 1.8$ and a column density of $10^{20.5}\pscm$. The red solid line shows the equivalent 1-to-1 line assuming a photon index of $\Gamma = 1.8$ and a column density of $10^{22.5}\pscm$. Black filled circles indicate sources with a measured column density of $<10^{20.5}\pscm$, dark red filled circles indicate sources with a measured column densities of $10^{20.5-22.5}\pscm$, red filled circles indicate sources with a measured column density of $10^{22.5-23}\pscm$, and orange filled circles indicate sources with a measured column density of $>10^{23}\pscm$. }
    \label{fig:bat_flux}
\end{figure}

The Swift-BAT and eRASS1 fluxes are plotted in Fig.~\ref{fig:bat_flux} for Piccinotti AGN (black, open squares), as well as the other sources in the BAT 70 month sample (circles), with colours indicating the different measured absorption column densities. The 70 month catalog is used in order to obtain the soft X-ray fluxes released in \citet{2017Ricci}. Open circles show the sources with non-detections in eROSITA which have measured column densities of N$_{\rm H}$<$10^{22.5}\pscm$. In order to compare the fluxes measured by different instruments and in different energy bands, we trace two solid lines; the black line shows the equivalent 1-to-1 line assuming a photon index of $\Gamma = 1.8$ and a column density of $10^{20.5}\pscm$, which are the approximately average parameters of the matched eROSITA and BAT sample. While this tends to pass through the approximate parameter space occupied by most sources, a number lie significantly above or below this line. 

Noticing that many of the sources which lie above the line show evidence for significant absorption, we add the second, red line which shows the equivalent 1-to-1 line assuming a photon index of $\Gamma = 1.8$ and a column density of $10^{22.5}\pscm$. Many sources with higher column densities are consistent with this line. This analysis demonstrates that fluxes measured between Swift and eROSITA are moderately in agreement, and does not reveal strong evidence for heavy pile-up in the hard X-ray detections of eROSITA.

\renewcommand{\arraystretch}{1.3}
\begin{table*}
\caption{Sample of Piccinotti AGN in the eROSITA-DE sky. Column (1) gives the common name, column (2) gives the eROSITA name, and column (3) gives the BAT name. Columns (4) and (5) give the optical RA and Dec, in degrees. Column (6) gives the eROSITA-derived observed $2.3-5\kev$ flux, and column (7) gives the observed flux from the swift BAT 70 month catalog derived from soft X-ray follow-up, e.g. with Swift-XRT or XMM-Newton. }
\label{tab:pic}
\resizebox{\textwidth}{!}{%
\begin{tabular}{lllccccc}
\hline
(1) & (2) & (3) & (4) & (5) & (6) & (7) & (8) \\
Name & eROSITA Name & BAT ID & RA & Dec & ML\_FLUX\_3 & BASS Flux & HEAO1 Flux \\
\hline
Fairall 9 & 1eRASS J012345.7-584819 & SWIFT J0123.9-5846 & 20.9408 & -58.8057 & 1.08E-11 & 2.73E-11 & 2.8E-11 \\
NGC 526a & 1eRASS J012354.3-350356 & SWIFT J0123.8-3504 & 20.9766 & -35.0654 & 1.14E-11 & 2.14E-11 &  5.5E-11 \\
ESO 198-024 & 1eRASS J023819.7-521133 & SWIFT J0238.2-5213 & 39.5821 & -52.1923 & 4.24E-12 & 1.29E-11 & 4.6E-11 \\
3C 120 & 1eRASS J043311.1+052114 & SWIFT J0433.0+0521 & 68.2962 & 5.3543 & 1.39E-11 & 3.69E-11 & 4.4E-11 \\
3A 0557-383 & 1eRASS J055802.0-382007 & SWIFT J0557.9-3822 & 89.5083 & -38.3346 & 3.91E-13 & 4.25E-11 & 3.0E-11 \\
H0917-074 & 1eRASS J092046.2-080323 & SWIFT J0920.8-0805 & 140.1927 & -8.0561 & 3.19E-12 & 1.29E-11 & 4.0E-11 \\
NGC 2992 & 1eRASS J094541.9-141935 & SWIFT J0945.6-1420 & 146.4252 & -14.3264 & 1.17E-11 & 7.30E-12 & 7.5E-11 \\
NGC 3227 & 1eRASS J102330.7+195153 & SWIFT J1023.5+1952 & 155.8774 & 19.8651 & 1.13E-11 & 3.69E-11 & 3.7E-11 \\
NGC 3783 & 1eRASS J113901.8-374418 & SWIFT J1139.0-3743 & 174.7572 & -37.7386 & 2.44E-11 & 7.69E-11 & 4.2E-11 \\
3C 273 & 1eRASS J122906.7+020308 & SWIFT J1229.1+0202 & 187.2779 & 2.0524 & 3.23E-11 & 1.18E-10 & 7.1E-11 \\
NGC 4593 & 1eRASS J123939.4-052038 & SWIFT J1239.6-0519 & 189.9143 & -5.3443 & 1.58E-12 & 3.90E-12 & 3.8E-11 \\
MCG -06-30-15 & 1eRASS J133553.8-341744 & SWIFT J1335.8-3416 & 203.9741 & -34.2956 & 1.45E-11 & 3.68E-11 & 4.6E-11 \\
IC 4329A & 1eRASS J134919.2-301834 & SWIFT J1349.3-3018 & 207.3303 & -30.3094 & 2.62E-11 & 9.63E-11 & 8.0E-11 \\
NGC 5506 & 1eRASS J141315.0-031227 & SWIFT J1413.2-0312 & 213.3119 & -3.2075 & 1.80E-11 & 9.27E-11 & 5.9E-11 \\
IRAS 18325-5926 & 1eRASS J183658.1-592408 & SWIFT J1836.9-5924 & 279.2429 & -59.4024 & 8.12E-13 & 2.45E-11 & 3.4E-11 \\
ESO 103-035 & 1eRASS J183819.7-652537 & SWIFT J1838.4-6524 & 279.5848 & -65.4276 & 2.58E-12 & 2.23E-11 & 3.0E-11 \\
H1846-786 & 1eRASS J184703.0-783148 & SWIFT J1848.0-7832 & 281.7618 & -78.5304 & 2.03E-12 & 6.20E-12 & 2.9E-11 \\
ESO 141-G055 & 1eRASS J192114.2-584013 & SWIFT J1921.1-5842 & 290.3090 & -58.6703 & 1.08E-11 & 2.17E-11 & 3.7E-11 \\
NGC 7213 & 1eRASS J220916.5-470959 & SWIFT J2209.4-4711 & 332.3177 & -47.1667 & 5.53E-12 & 1.18E-11 & 2.3E-11 \\
NGC 7582 & 1eRASS J231906.0-420648 & SWIFT J2318.4-4223 & 349.5979 & -42.3706 & 1.06E-12 & 2.30E-12 & 5.5E-11 \\
\hline
\hline
\end{tabular}
}
\end{table*}

\section{Variable point sources}
Boller \et (submitted) has analysed a sample of sources from the eRASS1 main catalog to check for evidence of variability within eRASS1. Due to the scanning pattern of eROSITA, this serves as a test of inter-eroday variability, corresponding to variability on 4-hour timescales. For each source, a light curve was produced \citep{Merloni2024}. Two variability tests are then performed: the maximum amplitude variability, defined as the span between the maximum and minimum values of the count rate, and the normalised excess variance (NEV), defined as the difference between the observed variance and the expected variance based on the errors \citep[see][Boller \et submitted]{2022Boller}. These variability tests are run in the $0.2-2.3\kev$ band, so while they do not overlap with our hard sample selection, they still provide an interesting look into the variability in the energy band where eROSITA is most sensitive. Variability in the hard X-ray is more difficult to disentangle from other events, including background flaring, pile-up and protons striking the detector. 

Matching the variable source catalog with the hard-band catalog, we find that 326 sources show evidence for variability. Out of these, 159 sources are within the LS10 area, and have good p\_any, suggesting these are secure counterparts. Of these 159, 40 sources have spectroscopic redshifts of $z>0.002$ and are thus confirmed to be extragalactic, and 109 are stars. Outside of the LS10 area, 125 sources have AGN matches (see appendix), five of which have a spectroscopic redshift of $z>0.002$ from Quaia. This gives a total of 45 variable AGN with a spec-z. These AGN are of particular interest for future follow-ups within later eRASSs or with other missions, to study both the short- and long-term variability of spectroscopic AGN. The low number of sources which are highly variable on short timescales also supports creating X-ray spectra from the full eRASS, rather than dividing by eroday. Such spectra will be further investigated in Section 8.2. Further details of the specific variability properties of these sources and a more detailed study of their variability is left for future work.

\section{X-ray spectral analysis}
\subsection{Spectroscopic AGN sample designation}
A key output of this classification work is a pure, hard X-ray selected sample of AGN detected with eROSITA. To ensure a very pure sample, we select only sources with reliable counterparts, and only those with spectroscopic redshifts, where $z>0.002$. More concretely, this sample is assembled from:

\begin{enumerate}
    \item All LS10 sources with a good counterpart and a spectroscopic redshift $z>0.002$, and removing sources with class\_beamed $ >0$ (1243 sources, 23 hard-only)
    \item Three AGN which were in the LS10 area but were missing counterparts, as listed in Table~\ref{tab:noctp} (three sources, zero hard-only)
    \item Sources from the Swift-BAT 105 month catalog crossmatch which satisfy the following criteria: eROSITA position outside of the LS10 area, no flags in the X-ray catalog, X-ray point source, BAT 105 month catalog Class of 40, 50, or 60 (corresponding to Seyfert 1, Seyfert 2, or LINER AGN, excluding beamed AGN and unknown type AGN) (82 sources, six hard-only)
\end{enumerate}

\begin{figure*}
    \centering
    \includegraphics[width=0.75\textwidth]{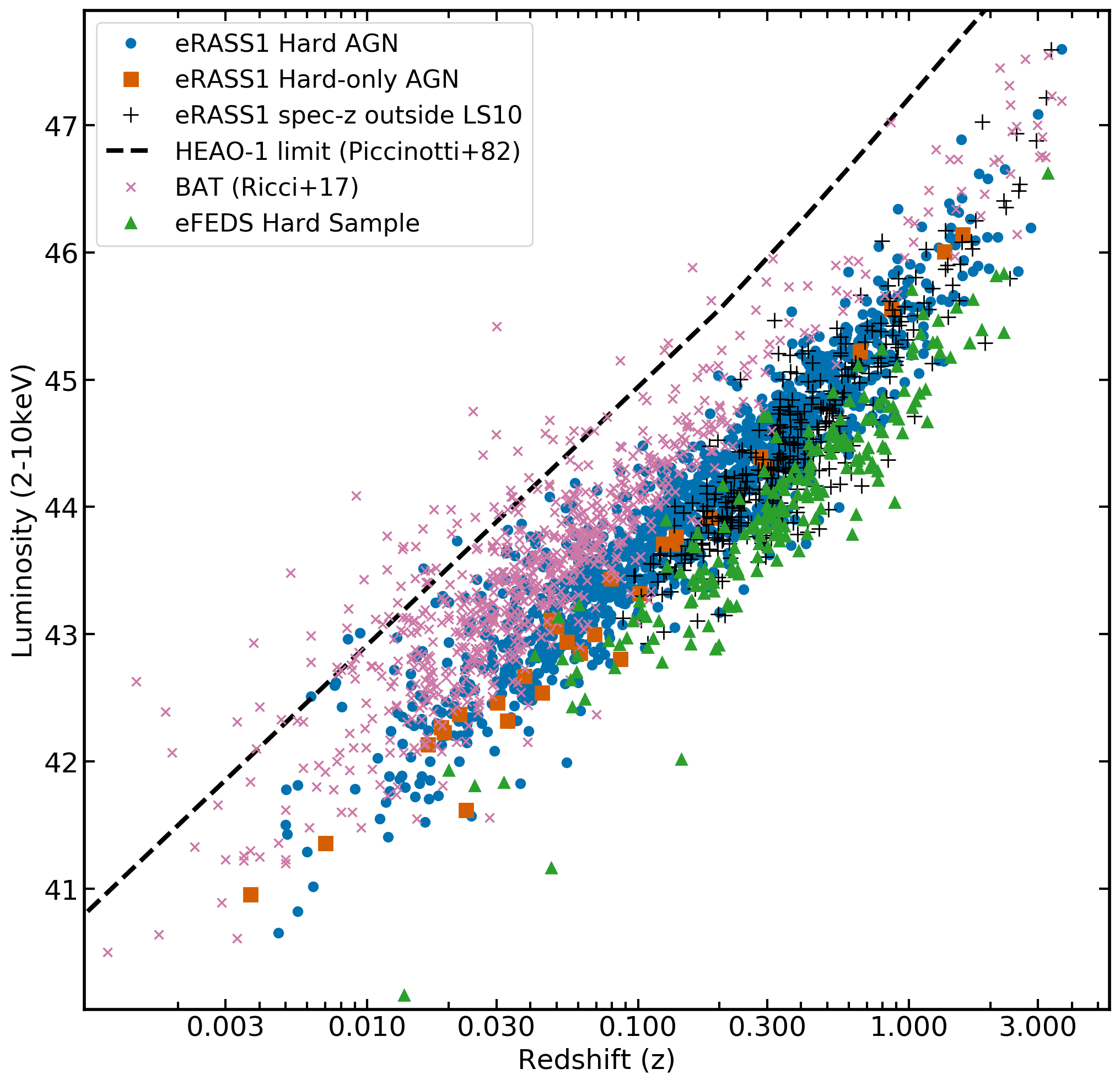}
    \caption{Hard X-ray luminosity versus redshift for various hard X-ray selected samples. eRASS1 AGN are shown as blue circles, eRASS1 hard-only AGN are shown with orange squares, Swift-BAT AGN from the 70 month catalog are shown as pink x, and the eFEDS hard sample are shown with green triangles. The flux limit of HEAO-1 is shown with a black dashed line. Sources outside LS10 with spec-z are shown as black crosses.}
    \label{fig:lz}
\end{figure*}

\noindent In total, this selects 1328 sources, of which 29 are hard-only, as can be seen in Table~\ref{tab:nums}. The luminosity-redshift distribution of these sources is shown in Fig.~\ref{fig:lz}. Sources detected in the BAT 70 month sample are also shown in pink, and sources detected in the eFEDS hard sample are indicated with green triangles. This sample is ideal for spectral analysis, since, as demonstrated by \citet{2023Waddell}, the presence of the photons above $2\kev$ will allow for excellent modelling of the contribution from the hot X-ray corona, as well as an investigation into soft X-ray emission and absorption. Furthermore, the eROSITA sample, in addition to being all-sky, is able to probe dimmer sources than the previous all-sky hard X-ray selected sample (HEAO-1, flux limit shown with the dashed black line) and than the BAT all-sky survey, and is also capable of finding many more sources at higher redshifts. The sample covers a broad redshift range of $0.003-3$, and an even broader hard X-ray luminosity range (estimated from the $2.3-5\kev$ band counts) of $10^{41}-10^{47}$\lumcgs. We caution that while only few eRASS1 sources appear above the HEAO1 flux limit, all 20 Piccinotti AGN in eROSITA\_DE are still detected, albeit at lower flux states.

Also shown here are the 339 sources with spectroscopic redshifts obtained from matching the sources outside the LS10 area to AGN catalogs, as described in Section 4 and the Appendix. These sources lie primarily within the Galactic plane and may be obscured by the local gas and dust, which may decrease the luminosities for these sources. A majority are also found at higher redshift, although examining the histogram in Fig.~\ref{fig:zhist}, we can see that the Quaia redshifts from the area in LS10 are similarly higher, so this may be due to the sensitivity and bandpass of the spectrograph onboard Gaia. It should also be noted here that some of the higher redshift sources may be blazars, and should be treated with caution.  

\subsection{X-ray spectral analysis}

Detailed X-ray spectral fitting of the entire spectroscopic redshift sample of 1328 sources is left for future work. Here we focus on the hard-only sources, of which 29 are not beamed and have spectroscopic redshifts $z>0.002$, 23 obtained from the \texttt{NWAY} LS10 match and six from the Swift-BAT match. This interesting subsample covers a large range in redshift $0.003<z<1.5$ and hard X-ray luminosity $10^{41}-10^{46}$\lumcgs, and are likely to be heavily obscured AGN. We fit each eROSITA X-ray spectrum with a torus model \citep[\texttt{UXClumpy};][]{2019Buchner}, using BXA \citep[Bayesian X-ray Analysis;][]{2014Buchnerbxa}. BXA uses nested sampling combined with XSPEC \cite[v. 12.12.1;][]{xspec} in order to explore the full model parameter space to best estimate parameters and enable model comparison. The model \texttt{UXClumpy} includes transmitted and reflected components with fluorescent lines obtained by assuming a clumpy torus with a specified opening angle and inclination. Since many of the eRASS1 spectra have very few counts, some model parameters are frozen to reasonable values obtained from \citet{2019Buchner}, including the width of the Gaussian distribution modelling the vertical extent of the torus cloud population (\texttt{TORsigma}, 20), and the viewing angle of the torus (\texttt{Theta\_inc}, 45$^{o}$). The photon index is given a Gaussian prior centered on 1.95 with a width of 0.15, and the column density varies uniformly between 20 and 26. 

\begin{figure}
    \centering
    \includegraphics[width=0.95\columnwidth]{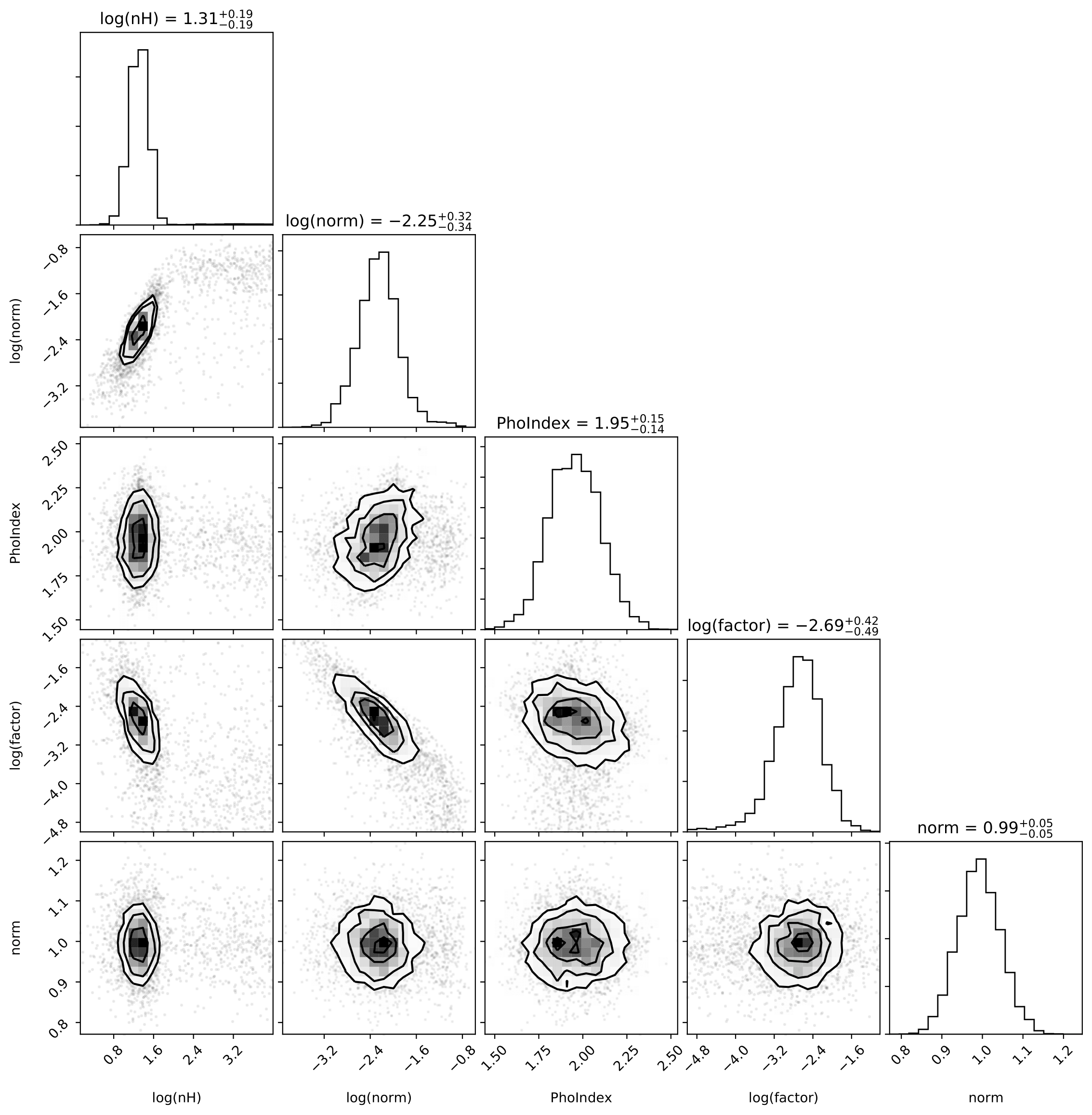}
    \caption{Corner plot resulting from applying the model to 1eRASS J151012.0-021454. The column log(nH) measures the torus column density in units of log(N$_{\rm H}$) $-22$, such that a value of $1$ corresponds to a column density of $10^{23}\pscm$, log(norm) is the normalisation, PhoIndex is the photon index, log(factor) is the renormalisation constant, and norm is the relative normalisation of the background model within the source field (should be $\sim1$). }
    \label{fig:corner}
\end{figure}

In order to model an additional soft population of X-ray photons, an additional power law is added, with photon index and normalisation linked to the \texttt{UXClumpy} component. This scattered power law is then renormalised by a constant factor which is left free to vary between 0.01 and 0.1. There are a number of interpretations for this soft X-ray emission; likely it represents a population of scattered photons which leak through the torus and pass into our line of sight \citep{2011Brightman,2014Liu,2016Furui,2019Buchner}, but soft X-ray emission can also be produced by other mechanisms such as supernova remnants, ultraluminous X-ray sources and X-ray binaries \citep[e.g. see discussion in][on Circinus]{2019Buchner}. within the host galaxy which is unresolved in X-rays. An example of the corner plot for source 1eRASS J151012.0-021454 (LEDA 54130) is shown in Fig.~\ref{fig:corner}. While some sources have double peaked solutions for the column density, others, as this one, are well-constrained, with little degeneracy between parameters. 

\begin{figure}
    \centering
    \includegraphics[width=0.95\columnwidth]{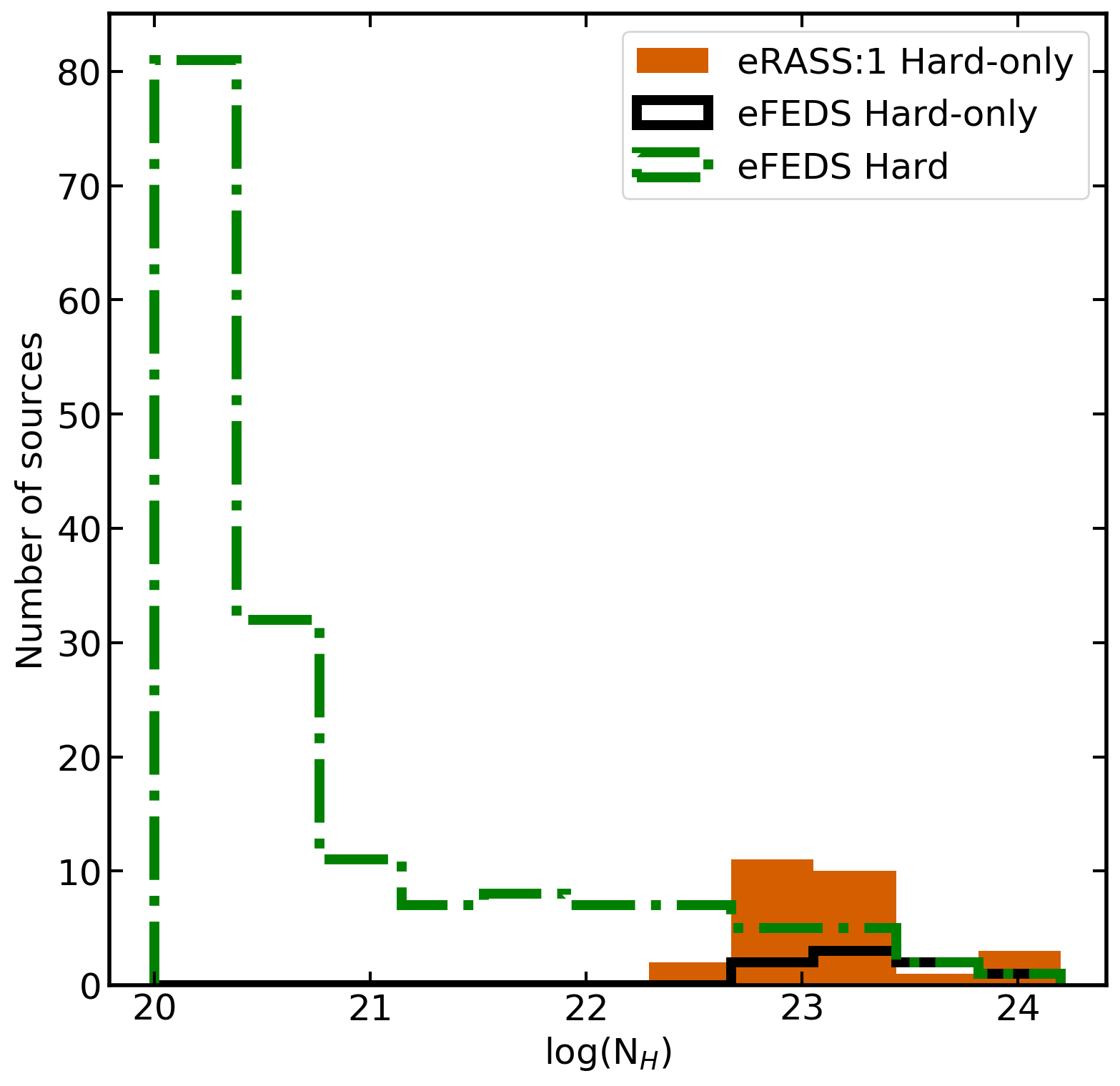}
    \caption{Distribution of column densities in the eFEDS hard (green dash-dot lines), eFEDS hard-only (black solid line), and eRASS1 Hard-only (orange filled histogram) samples. The column densities from eFEDS are shown for the AGN sample presented in Nandra \et (submitted), and the eRASS1 hard-only column densities are from the \texttt{UXClumpy} fits presented in this work. }
    \label{fig:nhefeds}
\end{figure}

The resulting column density distribution is shown in Fig.~\ref{fig:nhefeds}. The eRASS1 hard-only sources are shown in orange. A comparison is also given to the eFEDS sample, using spectral modelling as described in Nandra \et (submitted). The eFEDS hard sample is shown with a green dash-dot line, and the eFEDS hard-only sources are shown in black. While the hard sample as a whole contains many very bright, unobscured AGN, the hard-only sources recover a population of moderately obscured AGN, with column densities of $\sim10^{23-24}$\pscm. These column densities are high than what is often found in the discs of galaxies, so may originate in more obscured material closer to the central engine.

\begin{figure}
    \centering
    \includegraphics[width=0.95\columnwidth]{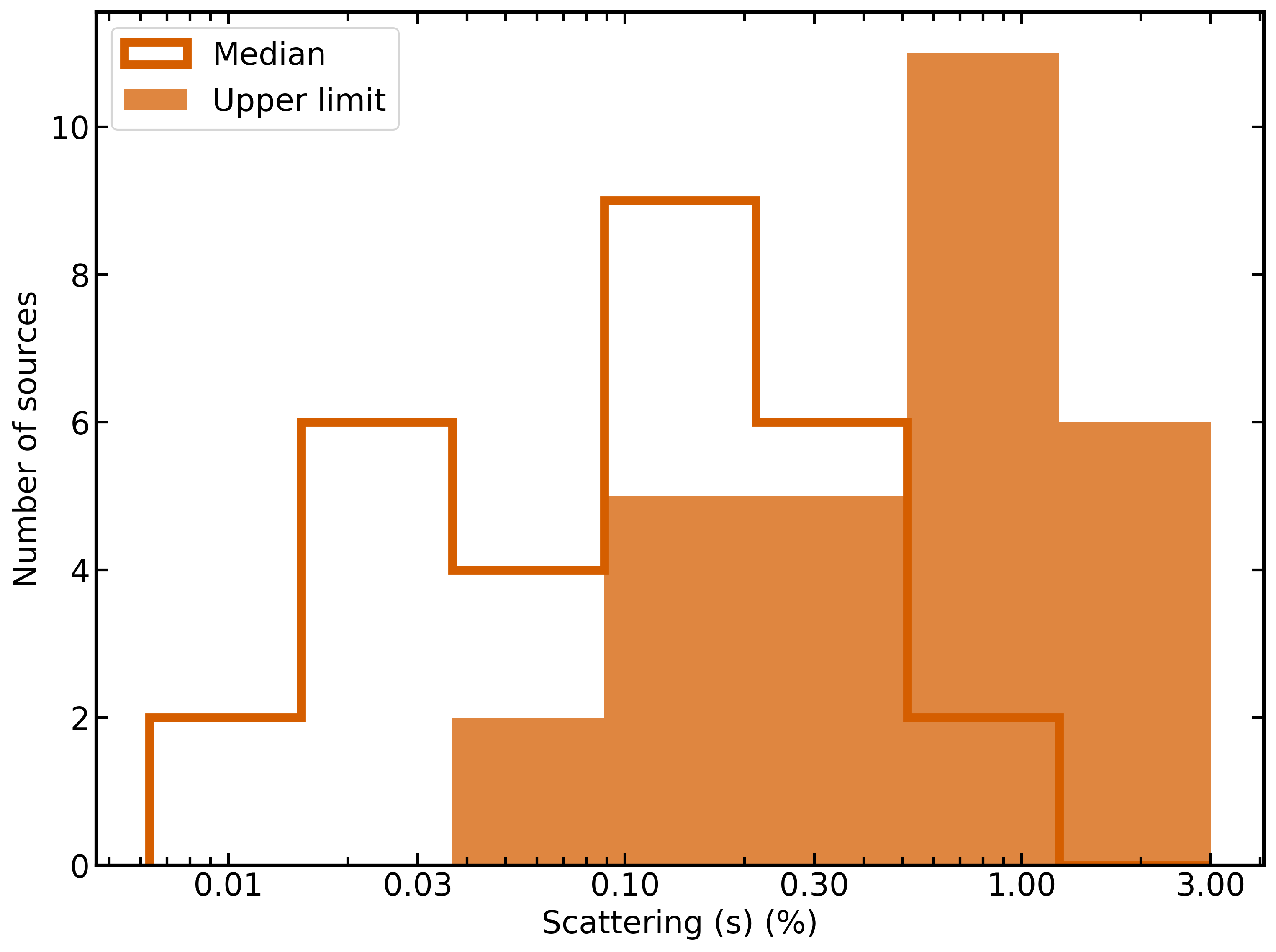}
    \includegraphics[width=0.95\columnwidth]{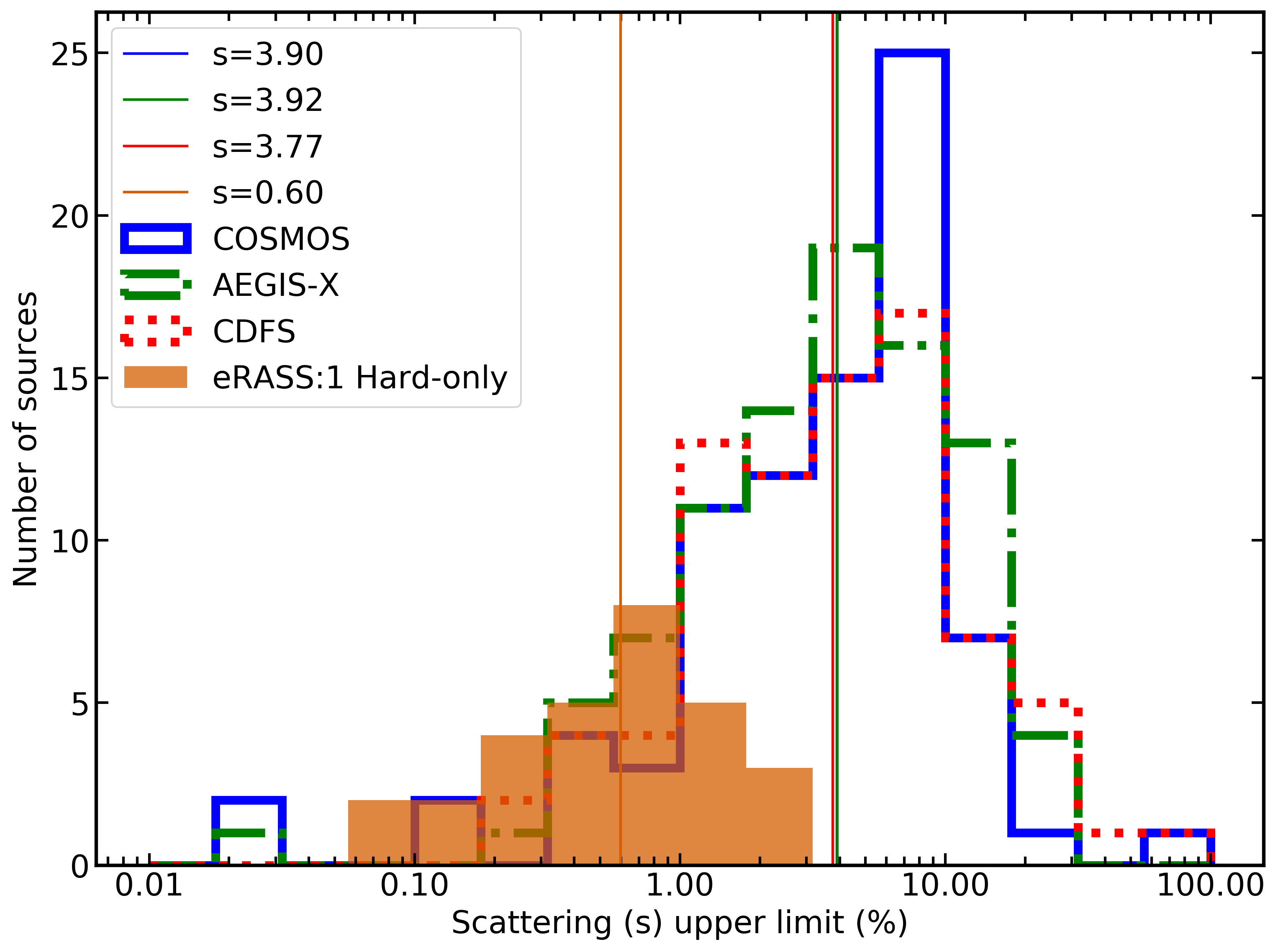}
    \caption{Top: Distribution of the posterior median of the scattering fraction (orange unfilled histogram) and upper limit of the scattering fraction (orange filled histogram), shown as a percentage, for the eRASS1 spectroscopic hard-only sample. Bottom: Distributions of the upper limits of the scattering fraction (shown as percentages) for this sample, in comparison to the best-fit Chandra results found by \citet{2014Brightman} The distributions measured from COSMOS, AEGIS-X and CDFS are shown as blue solid, green dash-dot and red dotted lines, respectively, all taken from \citet{2014Brightman} The eROSITA values are shown with an orange filled histogram. The median values are shown in vertical lines of corresponding colours, with almost exact overlap between the COSMOS, AEGIS-X and CDFS samples. }
    \label{fig:scatter}
\end{figure}

Also of interest here is the scattering fraction, which shows how many photons leak through the torus. We find that the upper limits for all 29 hard-only sources with spec-z of the scattering fractions are all $<3\%$, and the median values of the resulting posteriors are all $<1\%$, shown in the filled and unfilled histograms of Fig.~\ref{fig:scatter} (top), respectively. These correspond to $0.5-2\kev$ X-ray luminosities on the order of $10^{40}$\lumcgs. These values are lower than found in previous samples \citep{2014Brightman}. These are shown in Fig.~\ref{fig:scatter} (bottom), with the upper limits on the scattering fraction derived from COSMOS, AEGIS-X and CDFS shown compared to those from this work. Since eROSITA is extremely sensitive to soft X-ray photons, any small scattered emission should readily be detected, so by placing tight constraints on the detection threshold below $2.3\kev$, a more extreme source population is selected. There is some hypothesis that sources with low scattering fractions are associated with a buried AGN having a spherical torus which may arise in the late- or post-merger stages \citep[e.g.][]{2019Yamada}. However, looking at the LS10 images, we find many edge-on sources, in agreement with the very low b/a values found in Section 3.7 (see also Fig.~\ref{fig:ba}). Together, this may suggest that, particularly for the edge-on sources, some of the obscuration is associated with a $\sim$kpc scale obscurer in the host galaxy, versus solely in a parsec scale torus in the inner region. For the face-on sources, a uniform and spherical torus is preferred to reproduce the high level of obscuration and low scattering. 

\section{Discussion}

\subsection{Counterparts and associations}
From the first all-sky imaging survey in the $2-8\kev$ band, in the western galactic sky, \ero\ has detected 5466 sources with an estimated spurious fraction of about $10\%$. In this paper, we have identified and classified the 4895 clean, point-like sources present in the sample. For sources within the legacy survey DR10 area, we use \texttt{NWAY} to select counterparts for 2850 sources, and identify 2547 as having a secure counterpart. We then flag stars and beamed AGN and match with public spectroscopic redshift catalogs to finally select 1243 sources inside the LS10 are with $z>0.002$. 

One caveat to this analysis is that the priors in the \texttt{NWAY} run were constructed using a 4XMM and Chandra training sample which was created to match the eRASS1 main sample. The hard sample, however, is in general comprised of more bright, local AGN. Therefore, a majority of the AGN are brighter in optical and X-rays than the peak of the distributions from the training sample. The soft X-ray flux distributions are more comparable for the hard-only sources, but these may also have more extreme parameter values are they are a very distinct subclass of AGN. At the moment, there is not a large and well-defined enough training sample to select from, but using the counterparts from this sample to inform the training sample of future eROSITA all-sky surveys will help to improve counterparts and p\_any values for future works. Further, despite this caveat, it should be noted that there is an apparent 100\% overlap between the Swift BAT counterparts and the \texttt{NWAY} counterparts from this work, suggesting that \texttt{NWAY} is performing well for this sample. 

Examining the spectroscopic redshift distribution from Fig.~\ref{fig:zhist}, a majority of sources are very local, with a peak at $z\sim0.3$ which is in exact agreement with eFEDS (Nandra \et submitted). This is driven by the peaks of the sources with spectroscopic redshifts from NED, as well as from Quaia. Studying only the very secure spectroscopic redshifts from the compilation, the redshift peak is lower, at approximately $z\sim0.07-0.08$. The hard-only sources also have a surprisingly large spread of redshift values, from $z\sim0.03$ to $z>1$. Studying the luminosity-redshift values from Fig.~\ref{fig:lz}, we find that the all-sky survey selects many brighter sources than compared to eFEDS, but dimmer and higher redshift sources than BAT, providing a rich and unique sample well-suited for further exploration.

\subsection{Absorption in hard-only AGN}
A bona-fide hard-only sample can be produced by combining our hard-only selection with the information on the purity of the counterpart. Section 8.2 has demonstrated the power of X-ray spectral fitting 29 hard-only AGN with eROSITA, with remarkable results. Given the nature of the eROSITA survey strategy, the sources being fit have very few photons and low exposure times, which would typically suggest that there is little to be gained from spectral analysis. In fact, all sources have less than 20 counts, which was identified by \citet{2022Liu} as a cut-off below which little could be gained from parameter estimation and model comparison. However, spectral fitting is aided by a good understanding of the typical photon indices of eROSITA spectra \citep[e.g. from][Nandra \et submitted]{2022Liu} Crucially, for this work, fitting is enhanced not by what eROSITA detects, but what is not detected; simulations \citep{2022Seppi} have given an accurate estimation of the significance and spurious fraction of different detection thresholds. By selecting a very low threshold combined with the knowledge of eROSITA's extreme sensitivity to soft X-ray photons,  the hard-only sample is  very pure selection of sources with little to no soft X-ray emission (below $2.3\kev$). Correspondingly, the hard-only sample appears to consist of sources with high level of obscuration. 

Through spectral analysis, it is clear that not only are the column densities very high (with a median of log(N$_{\rm H}$) = 23.1), any percentage of photons scattered into the line of sight must also be extremely low, with a sample median of only 0.14\%. If this component is instead interpreted as emission from X-ray binaries or supernovae in addition to scattered photons, this implies that any scattered emission fraction is even lower than measured. As shown in Section 8.2, our values are much lower than what found from other samples \citep[e.g.][]{2014Brightman}. In \citet{2014Brightman}, data were obtained from Chandra, and the X-ray spectra were not analysed using \texttt{UXClumpy}, but rather with \texttt{BNtorus} \citep{2011Brightman}, using an older version of this code which allowed soft X-ray leakage, making the spectral shape extremely similar to \texttt{UXClumpy}. Therefore, comparison between measured values is justified. The differences do not suggest an intrinsic difference between AGN detected between the two missions, and spectral modelling of the full eRASS1 hard sample would likely return a more similar overall scattering fraction distribution to the Chandra samples from \citet{2014Brightman}. Rather, this analysis demonstrates that the hard-only sample is a highly unique sub-sample whose properties may be of particular interest, and that this is distinct from other surveys and samples.  

In Section 8.2, it was also discussed that the small recovered scattering fractions may imply that the obscuration is associated with a $\sim$kpc scale obscurer on larger scales than the typical X-ray torus. This result was further supported by the edge-on galaxies in the images, and also by the lower b/a values measured in LS10. This may contrast to what is suggested in other works, where it is hypothesized that the lower scattering fractions are more indicative of a spherical torus \citep[e.g.][]{2019Yamada}. For sources that are face-on, a uniform and spherical torus is likely the preferred interpretation. This is also contrary to many works which suggest that the torus is patchy and not fully obscuring \citep[e.g.][]{1999Maiolino,2008Nenkova,2014Markowitz,2015Aird,2015Buchner,2019Buchner}. The hard-only sample therefore allows us to probe a unique and extreme parameter space of torus structure. This interesting result could be studied in more detail using broad-band X-ray data, with a deeper observation of the soft X-ray to search for a scattered component or for time variability, and of the hard X-ray to model the shape and parameters of the obscurer and underlying power law.

\section{Conclusions}
eRASS1 is the first all-sky imaging survey above $2\kev$, with the eROSITA-DE half-sky being publicly released. The hard X-ray selection decreases the bias against obscured sources compared
to e.g. the eRASS1 main sample. The sample presents an extraordinary improvement over the Piccinotti AGN sample, and allows us to probe deeper in flux and redshift than e.g. the Swift-BAT all-sky survey. In addition to the counterparts and multiwavelength information provided, we identify a population of heavily obscured AGN, in the hard-only sample, which lack detection in eROSITAs most sensitive soft energy, but are detected in the hard band. Through X-ray spectral analysis of these, combined with the images and optical analysis, it was hypothesized that the obscuration of these sources could be associated with both a spherical torus and/or the galactic disc. Follow-up observations of these sources with other X-ray missions will help to more accurately model the X-ray spectra. Continued optical spectroscopic follow-up of these sources (e.g. with SDSS-V), and indeed of the entire hard sample, will help to identify more AGN, and hard-only AGN. Finally, combining these results with future X-ray all-sky surveys (including eRASS:2, eRASS:3 and eRASS:4) will help to further secure the source catalog and counterparts, shining a new light on the nature of X-ray sources above $2\kev$.

\begin{acknowledgements}
This work is based on data from eROSITA, the soft X-ray instrument aboard SRG, a joint Russian-German science mission supported by the Russian Space Agency (Roskosmos), in the interests of the Russian Academy of Sciences represented by its Space Research Institute (IKI), and the Deutsches Zentrum für Luft- und Raumfahrt (DLR). The SRG spacecraft was built by Lavochkin Association (NPOL) and its subcontractors, and is operated by NPOL with support from the Max Planck Institute for Extraterrestrial Physics (MPE).

The development and construction of the eROSITA X-ray instrument was led by MPE, with contributions from the Dr. Karl Remeis Observatory Bamberg \& ECAP (FAU Erlangen-Nuernberg), the University of Hamburg Observatory, the Leibniz Institute for Astrophysics Potsdam (AIP), and the Institute for Astronomy and Astrophysics of the University of Tübingen, with the support of DLR and the Max Planck Society. The Argelander Institute for Astronomy of the University of Bonn and the Ludwig Maximilians Universität Munich also participated in the science preparation for eROSITA.

The eROSITA data shown here were processed using the eSASS/NRTA software system developed by the German eROSITA consortium.

The Legacy Surveys consist of three individual and complementary projects: the Dark Energy Camera Legacy Survey (DECaLS; Proposal ID \#2014B-0404; PIs: David Schlegel and Arjun Dey), the Beijing-Arizona Sky Survey (BASS; NOAO Prop. ID \#2015A-0801; PIs: Zhou Xu and Xiaohui Fan), and the Mayall z-band Legacy Survey (MzLS; Prop. ID \#2016A-0453; PI: Arjun Dey). DECaLS, BASS and MzLS together include data obtained, respectively, at the Blanco telescope, Cerro Tololo Inter-American Observatory, NSF’s NOIRLab; the Bok telescope, Steward Observatory, University of Arizona; and the Mayall telescope, Kitt Peak National Observatory, NOIRLab. Pipeline processing and analyses of the data were supported by NOIRLab and the Lawrence Berkeley National Laboratory (LBNL). The Legacy Surveys project is honored to be permitted to conduct astronomical research on Iolkam Du’ag (Kitt Peak), a mountain with particular significance to the Tohono O’odham Nation.

NOIRLab is operated by the Association of Universities for Research in Astronomy (AURA) under a cooperative agreement with the National Science Foundation. LBNL is managed by the Regents of the University of California under contract to the U.S. Department of Energy.

This project used data obtained with the Dark Energy Camera (DECam), which was constructed by the Dark Energy Survey (DES) collaboration. Funding for the DES Projects has been provided by the U.S. Department of Energy, the U.S. National Science Foundation, the Ministry of Science and Education of Spain, the Science and Technology Facilities Council of the United Kingdom, the Higher Education Funding Council for England, the National Center for Supercomputing Applications at the University of Illinois at Urbana-Champaign, the Kavli Institute of Cosmological Physics at the University of Chicago, Center for Cosmology and Astro-Particle Physics at the Ohio State University, the Mitchell Institute for Fundamental Physics and Astronomy at Texas A\&M University, Financiadora de Estudos e Projetos, Fundacao Carlos Chagas Filho de Amparo, Financiadora de Estudos e Projetos, Fundacao Carlos Chagas Filho de Amparo a Pesquisa do Estado do Rio de Janeiro, Conselho Nacional de Desenvolvimento Cientifico e Tecnologico and the Ministerio da Ciencia, Tecnologia e Inovacao, the Deutsche Forschungsgemeinschaft and the Collaborating Institutions in the Dark Energy Survey. The Collaborating Institutions are Argonne National Laboratory, the University of California at Santa Cruz, the University of Cambridge, Centro de Investigaciones Energeticas, Medioambientales y Tecnologicas-Madrid, the University of Chicago, University College London, the DES-Brazil Consortium, the University of Edinburgh, the Eidgenossische Technische Hochschule (ETH) Zurich, Fermi National Accelerator Laboratory, the University of Illinois at Urbana-Champaign, the Institut de Ciencies de l’Espai (IEEC/CSIC), the Institut de Fisica d’Altes Energies, Lawrence Berkeley National Laboratory, the Ludwig Maximilians Universitat Munchen and the associated Excellence Cluster Universe, the University of Michigan, NSF’s NOIRLab, the University of Nottingham, the Ohio State University, the University of Pennsylvania, the University of Portsmouth, SLAC National Accelerator Laboratory, Stanford University, the University of Sussex, and Texas A\&M University.

BASS is a key project of the Telescope Access Program (TAP), which has been funded by the National Astronomical Observatories of China, the Chinese Academy of Sciences (the Strategic Priority Research Program “The Emergence of Cosmological Structures” Grant \# XDB09000000), and the Special Fund for Astronomy from the Ministry of Finance. The BASS is also supported by the External Cooperation Program of Chinese Academy of Sciences (Grant \# 114A11KYSB20160057), and Chinese National Natural Science Foundation (Grant \# 12120101003, \# 11433005).

The Legacy Survey team makes use of data products from the Near-Earth Object Wide-field Infrared Survey Explorer (NEOWISE), which is a project of the Jet Propulsion Laboratory/California Institute of Technology. NEOWISE is funded by the National Aeronautics and Space Administration.

The Legacy Surveys imaging of the DESI footprint is supported by the Director, Office of Science, Office of High Energy Physics of the U.S. Department of Energy under Contract No. DE-AC02-05CH1123, by the National Energy Research Scientific Computing Center, a DOE Office of Science User Facility under the same contract; and by the U.S. National Science Foundation, Division of Astronomical Sciences under Contract No. AST-0950945 to NOAO.

MB and BM are supported by the European Union's Innovative Training  Network (ITN) "BiD4BEST", funded by the  Marie  Sklodowska-Curie Actions in Horizon 2020 (GA N. 860744). ZI and JW acknowledge support by the Deutsche Forschungsgemeinschaft (DFG, German Research Foundation) under Germany’s Excellence Strategy - EXC-2094 - 390783311. 

This research has made use of the SIMBAD database, operated at CDS, Strasbourg, France. The following software packages/tools were also used: astropy \citep{2022astropy}, matplotlib \citep{2007Hunter}, numpy \citep{2020numpy}, topcat \citep{2005Topcat}, xspec \citep{xspec} 

\end{acknowledgements}

%
\bibliographystyle{aa} 
\bibliography{erass1_hard_arxiv} 
%

\begin{appendix} 
\section{LS10 \texttt{NWAY} Catalog description}
The following describes each column in the released catalog of counterparts obtained by matching with \texttt{NWAY} within the LS10 \citep{2019Dey} area, as described in Section 3. Columns $1-109$ are the eROSITA X-ray catalog data and are copied directly from the catalog released by \citet{Merloni2024}, and all columns are described in that paper. All columns indicated as "(from LS10)" are copied from the Legacy Survey DR10 catalogs.

\noindent \textbf{InAllLS10:} True if the source is within the LS10 MOC \\
\textbf{LS10\_RELEASE:} Release column denotes the camera and filter set used (from LS10) \\
\textbf{LS10\_BRICKID:} Brick ID column (from LS10) \\
\textbf{LS10\_OBJID:} Object \\
\textbf{LS10\_RA:} LS10 right ascension (J2000) (from LS10) \\
\textbf{LS10\_DEC:} LS10 declination (J2000) (from LS10) \\
\textbf{LS10\_Xray\_proba:} Probability that the LS10 source is an X-ray emitter\\
\textbf{Separation\_max:} Separation between the LS10 source and the eROSITA position\\
\textbf{p\_any:} Probability that the X-ray source has a counterpart \\
\textbf{p\_i:} Relative probability of the LS10 counterpart \\
\textbf{TYPE:} type column (from LS10), indicating whether the optical source is point-like (PSF) or extended (not PSF)  \\
\textbf{FLUX\_G:} g band flux (from LS10) \\
\textbf{FLUX\_R:} r band flux (from LS10) \\
\textbf{FLUX\_I:} i band flux (from LS10) \\
\textbf{FLUX\_Z:} z band flux (from LS10) \\
\textbf{FLUX\_W1:} W1 (unWISE) flux (from LS10) \\
\textbf{FLUX\_W2:} W2 (unWISE) flux (from LS10) \\
\textbf{FLUX\_IVAR\_G:} Inverse variance of G band flux (from LS10) \\
\textbf{FLUX\_IVAR\_R:} Inverse variance of R band flux (from LS10) \\
\textbf{FLUX\_IVAR\_I:} Inverse variance of I band flux (from LS10) \\
\textbf{FLUX\_IVAR\_Z:} Inverse variance of Z band flux (from LS10)  \\
\textbf{FLUX\_IVAR\_W1:} Inverse variance of W1 flux (from LS10) \\
\textbf{FLUX\_IVAR\_W2:} Inverse variance of W2 flux (from LS10)\\
\textbf{SHAPE\_R:} Half-light radius of sources where the type is not PSF (from LS10) \\
\textbf{SHAPE\_R\_IVAR:} Inverse variance on the half-light radius (from LS10) \\
\textbf{SHAPE\_E1:} Ellipticity component 1 for sources where the type is not PSF (from LS10) \\
\textbf{SHAPE\_E1\_IVAR:} Inverse variance of ellipticity component 1 (from LS10) \\
\textbf{SHAPE\_E2:} Ellipticity component 2 (complex component) for sources where the type is not PSF (from LS10) \\
\textbf{SHAPE\_E2\_IVAR:} Inverse variance of ellipticity component 2 (from LS10) \\
\textbf{REF\_CAT:} Reference catalog taken from LS10; Tycho-2, Gaia EDR3 or SGA (from LS10) \\
\textbf{REF\_ID:} ID from reference catalog (from LS10)\\
\textbf{PARALLAX:} Reference catalog parallax (from LS10) \\
\textbf{PARALLAX\_IVAR:} Inverse variance of parallax (from LS10) \\
\textbf{redshift\_ZSPEC:} Redshift from spectroscopic redshift compilation \\
\textbf{redshift\_err\_ZSPEC:} Error on redshift from compilation\\
\textbf{ref\_ZSPEC:} Reference from which spectroscopic redshift was retrieved \\
\textbf{redshift\_NED:} NED spectroscopic redshift \\
\textbf{redshift\_quaia:} Redshift from Quaia \\
\textbf{redshift\_quaia\_err:} Error on redshift from Quaia \\
\textbf{dered\_mag\_g:} g band flux converted to magnitude, corrected for Milky Way (MW) reddening \\
\textbf{dered\_mag\_r:} r band flux converted to magnitude, corrected for MW reddening\\
\textbf{dered\_mag\_z:} z band flux converted to magnitude, corrected for MW reddening \\
\textbf{dered\_mag\_w1:} W1 band flux converted to magnitude, corrected for MW reddening \\
\textbf{dered\_mag\_w2:} W2 band flux converted to magnitude, corrected for MW reddening \\
\textbf{ML\_FLUX\_12:} Summed X-ray flux from band 1 ($0.2-0.6\kev$) and 2 ($0.6-2.3\kev$)\\
\textbf{Gaia\_moving\_5sigma:} Whether or not the parallax is significant at or above the 5 sigma level \\
\textbf{r\_mag:} r magnitude, not corrected for Galactic reddening \\
\textbf{SWIFTBAT\_ID:} ID from SWIFT \citep{2018Oh} \\
\textbf{SWIFT\_SWIFTBAT70:} ID from SWIFT BAT 70 month catalog \citep{2017Ricci} \\
\textbf{hardonly:} Flag for whether or not the source is hard-only (1 if true, 0 if false) \\
\textbf{redshift\_source:} Origin of the redshift; 5 means redshift catalog, 2 means NED, 1 means Quaia, 0 means no redshift\\
\textbf{redshift\_best:} Final, best spectroscopic redshift as described in the text \\
\textbf{SIMBAD\_known\_galactic:} If the source is classified as Galactic in SIMBAD \\
\textbf{class\_beamed:} Flag for if the source appears to be beamed; 3 means it is in SIMBAD, as some form of beamed AGN, 2 means it is in CRATES, 1 means it is in BZCAT, and 0 means it is not in any of these catalogs  \\

\section{Catalog of AGN Outside LS10}
The following describes each column in the released catalog of counterparts obtained by performing a positional match for the X-ray sources outside of the LS10 area, as described in Section 4. Only the sources with both hard and soft X-ray detections are used, leaving 1710 X-ray sources. Three catalogs are used to identify candidate AGN; the Gaia/Quaia AGN sample \citep{2023Storey}, the Gaia/unWISE catalog \citep{2019Shu} and the Million Quasar Catalog \citep[MilliQSO;][]{2023Flesch}. Since the area covers the Galactic plane and includes many stars, matching only to AGN catalogs simplifies the matching process and still provides insight into the redshifts of sources outside the LS20 area. In order to perform the positional match, a five arcsecond match radius is used, comparing the eROSITA X-ray positions to the positions from each catalog. Since the positional errors of the not hard-only sources are small, it is reasonable to assume that the X-ray emission is likely to originate in the nearest AGN, but future work using \texttt{NWAY} to match with WISE and Gaia catalogs will help to better inform these counterparts. In total, 692 matches are recovered, 339 of which have a spectroscopic redshift from Quaia.  Columns $1-109$ are the eROSITA X-ray catalog data and are copied directly from the catalog released by \citet{Merloni2024}, and all columns are described in that paper. 

\noindent \textbf{Quaia\_source\_id:} Closest match Quaia source ID \\
\textbf{redshift\_quaia:} Quaia redshift \\
\textbf{redshift\_quaia\_err:} Error on Quaia redshift \\
\textbf{Gaia\_unwise\_sourceid:} Closest match Gaia/unwise source ID \\
\textbf{Gaia\_unwise\_PHOT\_Z:} Gaia/unwise Photo-z \\
\textbf{Milliqso\_NAME:} Milliquas catalog source name \\
\textbf{Milliqso\_Z:} Milliquas catalog redshift \\
\textbf{Milliqso\_ZCITE:} Milliquas catalog redshift reference \\
\textbf{SWIFTBAT\_ID:} ID from SWIFT \citep{2018Oh} \\

\end{appendix}

\end{document}